\documentclass[8pt]{article}
\usepackage[a4paper, total={6.5in, 9in}]{geometry}
\usepackage{amsmath,amssymb,amsthm}
\usepackage{amsfonts}
\usepackage{mathtools}
\usepackage{breqn}
\usepackage{bm}
\usepackage{graphicx}
\usepackage{epstopdf}
\usepackage{subcaption}
\usepackage{authblk}
\usepackage{color,soul}
\usepackage[english]{babel}
\usepackage{stackengine}
\usepackage{hyperref}
\usepackage{array}
\usepackage{tikz}
\usepackage{tikz-3dplot}
\usetikzlibrary{arrows.meta, positioning, shapes.geometric}
\usepackage{algorithm}
\usepackage{algpseudocode}
\usepackage{cite}
\usepackage{lineno}
 
\usepackage{bigints}
\usepackage[title,titletoc,toc]{appendix}
\usepackage{etoolbox}
\appto\appendices{\counterwithin{equation}{section}}
\usepackage{epsfig}

\hypersetup{
    colorlinks=true,
    linkcolor=blue,
    filecolor=blue,      
    urlcolor=blue,
    citecolor=blue
}

\newcommand{\bs}{\boldsymbol}
\newcommand{\tb}{\textbf}

\newcommand\groupequation[2][30pt]{%
  \setbox0=\hbox{$\displaystyle#2$}%
  \stackengine{0pt}{\copy0}{%
    \makebox[\linewidth]{\hfill$\left.\rule{0pt}{\ht0}\right\}$\kern#1}}
    {O}{c}{F}{T}{L}
}

\begin{document}
\title{A novel Grand-Potential Phase-Field Lattice-Boltzmann model for multi-phase solidification with convection.}

\author[1]  {Chirantandip Mahanta   \thanks{chirantandi5@iisc.ac.in (corresponding author)}}
\author[1,3]{Sanjeev Kumar          \thanks{sanjeev.kumar@austin.utexas.edu}}
\author[2]  {Gandham Phanikumar     \thanks{gphani@iitm.ac.in}}
\author[1]  {Abhik N. Choudhury     \thanks{abhiknc@iisc.ac.in}}
\affil[1]{\small Materials Modelling Lab, Department of Materials Engineering, Indian Institute of Science.}
\affil[2]{\small Department of Metallurgical and Materials Engineering, Indian Institute of Technology Madras.}
\affil[3]{\small Department of Aerospace Engineering and Engineering Mechanics, The University of Texas at Austin.}
\date{}
\maketitle
\begin{abstract}   


Melt convection plays a critical role in microstructure evolution during alloy solidification, yet accurately capturing its interaction with moving solid–liquid interfaces remains a significant computational challenge, particularly in multi-phase, multi-component systems. In this work, we develop a computational framework that couples a Grand-Potential phase-field model with the Lattice Boltzmann method (LBM) to simulate convection-driven solidification within a unified and thermodynamically consistent formulation. The proposed approach rigorously enforces no-slip conditions at evolving solid–liquid interfaces, while fluid transport is solved using the standard single-relaxation-time Bhatnagar–Gross–Krook collision operator. The framework provides an efficient and robust methodology for resolving the coupled evolution of interfaces, solute transport, and fluid flow.

The versatility of the proposed framework is demonstrated through simulations of dendritic and eutectic solidification under natural convection. The results show that convection significantly modifies solute segregation, destabilizes growth fronts, and induces oscillatory growth modes in both systems. These examples illustrate the capability of the proposed method to capture complex flow-induced morphological instabilities and provide new insights into the mechanisms governing convection-driven microstructure evolution in alloy solidification.
\end{abstract}


\section{Introduction}
\label{sec:intro}

Natural convection is an inevitable phenomenon arising from the density changes in the liquid phase during solidification processes. It not only alters the morphology of the growing solid-liquid interface but also causes the formation of defects like freckles and porosity\cite{dantzigBook}. Experiments have conclusively shown the influence of convection on dendritic features like primary and secondary arms spacings, tip radius, growth direction etc.\cite{Spinelli2004,Liu2006,Shevchenko2015} Phenomenon like columnar-to-equiaxed transitions, fragmentation of dendrites and other instabilities have also been directly linked to natural convection\cite{Ruvalcaba2007, Reinhart2013, Shevchenko2013}. Additionally, the role of convection in influencing the morphology of multi-phase solidification remains largely elusive. Therefore a comprehensive understanding of the dynamics of fluid-solid interactions is necessary for producing desired microstructures.


Computational modeling techniques provide an edge over experiments in this regard, as they offer a precise control on the physics and the environment to study how the microstructure responds to the flow field. Over time, different approaches have been used to study the influence of convection on a growing dendritic morphology. Notable among them are the Cellular-Automaton Models\cite{Reinhart2013, Zhang2022, Zhang2024}, the Mesoscopic Envelope model\cite{Viardin2020}, the Needle-Network Model\cite{Isensee2022}, and the Phase-Field (PF) Models. The PF models coupled with Navier-Stokes have a distinct advantage of accuracy and robustness over the others\cite{Viardin2020}. But rather than solving the Navier-Stokes equations directly, researchers have preferred using the Lattice-Boltzmann formalism due its its simplicity and ease of parallelization. Thus there has been a growing popularity of the Phase-Field-Lattice-Boltzmann (PF-LBM) methods in investigating the influence of convection in solidification processes. Many such earlier investigations have discounted natural convection and focused primarily on the influence of forced convection\cite{Selzer2009,Rojas2015,Sakane2017}. Recently, significant progress has been made by investigating the solidification of binary alloy systems and phenomenon like dendritic collision, sedimentation, columnar to equiaxed transitions, growth stability etc. due to the development of large scale massively parallel PF-LBM codes\cite{Takaki2017,Takaki2018,Ratkai2019,Takaki2020}. Additionally, the PF-LBM models have been extended to multi-component alloy systems, and dendritic structures of CMSX-4 have been shown to exhibit the typical convection driven effects\cite{Yang2019}. 

Convection effects on multi-phase systems eutectics however, have not been investigated as extensively. The only known work (to the authors best knowledge) is a study of the binary Al-Cu eutectic with a PF-LBM model, under isothermal conditions, showing the effects of convection on solute transport and steady-state growth behavior\cite{Zhang2018_eu,Zhang2019_eu}. A detailed review of the PF-LBM methods is presented by Samanta et. al.\cite{Samanta2022}.

In essence, multi-phase multi-component systems have remained largely un-investigated due to their computational complexity and, the necessity for a general framework for multi-component and multi-phase systems and one that accounts for true thermodynamics (from databases) remained.

In this paper we have developed a Grand-Potential Phase-Field model coupled with the Lattice-Boltzmann method, that is capable of solving a generic multi-phase multi-component alloy system and can account for accurate alloy thermodynamics (from databases). We have incorporated three forces that influence fluid flow namely, the drag force, the buoyancy force and the shrinkage force in the Lattice-Boltzmann equation for completeness of the physics involved. Equipped with this model, we investigate the nature and emergence of morphological instabilities in dendritic and eutectic solidification processes, and the underlying mechanisms of the flow microstructure interactions.

\section{The Phase-Field-Lattice-Boltzmann Model}
\label{sec:phase-field_LBM}

This section presents the coupling between the Grand Potential based Phase-Field model and the Lattice-Boltzmann method.

\subsection{Phase-Field Model}
\label{sec:phase-field}

In the Grand-Potential Phase-Field (GPPF) formulation\cite{choudhury2012grand}, for a system with $N$ phases and $K$ species, the governing equations for the phase-fields \{$\phi_{\alpha}$\} and chemical potentials \{$\mu_i$\} are given as:

\begin{equation}
  \tau \epsilon \frac{\partial \phi_\alpha}{\partial t} = \epsilon \bigg[\nabla \cdot \frac{\partial a(\bs{\phi}, \nabla \bs{\phi})}{\partial \nabla \phi_\alpha} - \frac{\partial a(\bs{\phi}, \nabla \bs{\phi})}{\partial \phi_\alpha} \bigg] - \frac{1}{\epsilon} \frac{\partial W (\bs{\phi})}{\partial \phi_\alpha} - \frac{\partial \Psi (\tilde{\bs{\mu}}, T,\bs{\phi})}{\partial \phi_\alpha} -  \Lambda,
  \label{eq:PhfGrand}
\end{equation}
\begin{eqnarray}
\nonumber
\left\{ \frac{\partial \Tilde{\mu}_i}{\partial t} \right\} 
    =   \left[\sum_{\alpha = 1}^N h_\alpha(\bs{\phi}) \frac{\partial C_i^\alpha(\Tilde{\bs{\mu}},T)}{\partial \Tilde{\mu}_j} \right]^{-1}_{ij}  
        \Bigg\{ 
            \nabla \cdot \sum_{j=1}^{K-1} M_{ij}(\bs{\phi}) \nabla \Tilde{\mu}_j 
            - \sum_{\alpha=1}^N C_i^\alpha(\Tilde{\bs{\mu}},T) \frac{\partial h_\alpha(\bs{\phi})}{\partial t} \\
            -\sum_{\alpha = 1}^N \tb{U} \cdot \nabla {C}_i^\alpha
            -\frac{\partial T}{\partial t} \sum_{\alpha=1}^N \left(\frac{\partial C_i^\alpha(\Tilde{\bs{\mu}}^m,T)}{\partial T}\right)_{\Tilde{\mu}} h_\alpha(\bs{\phi})
        \Bigg\},
\label{eq:conc}
\end{eqnarray}
here we have modified the standard GPPF model by adding an advection term to the composition evolution part of Equation.\ref{eq:conc}. We will assume the solid phases to be stationary and consider mass advection only in the liquid phase. Hence the velocity field $\bs{U}$ is assumed to be $\bf{0}$ in all the solid phases. In the liquid phase $\bf{U}$ will be solved using the Lattice-Boltzmann method described in the next section.


The term $\Lambda$ in Eq. \ref{eq:PhfGrand} denotes the Lagrange parameter for the constraint $\sum_{\alpha = 1}^N \phi_\alpha = 1$. 
Anisotropy of the interfaces is embedded in the gradient energy density function  $a(\bs{\phi}, \nabla \bs{\phi})$ and $W$ is the surface potential density as double obstacle function, given as:
\begin{equation}
W(\bs{\phi}) = \frac{16}{\pi^2} \sum^{N,N}_{\alpha, \beta=1\; 
 (\alpha<\beta)} \gamma_{\alpha \beta} \phi_\alpha \phi_\beta ,
 \label{eq:Wpot}
\end{equation}
we set $W(\bs{\phi}) = \infty$, if $\bs{\phi}$ is not on the Gibbs simplex given as, $G=\{ \bs{\phi} \in \mathbb{R}^N : \sum_\alpha\phi_\alpha =1, \: \forall \: \phi_\alpha \geq 0\}$. 

We have retained the standard GPPF method of coupling with arbitrary thermodynamic systems by reading in data from the CALPHAD databases. The details of which are in Ref\cite{Choudhury2015}. This coupling enables us to solve for arbitrary multi-phase multi-component systems with ease.

\subsection{Lattice-Boltzmann Method}
\label{subsec:LBM}

At the core of the Lattice-Boltzmann method\cite{LBM_Chen1998} lies the discrete-velocity distribution function $f_k$, along the direction $k$, which along with the BGK collision operator and a force term has the form:
\begin{equation}
    f_k \left( \bs{x} + \bs{c}_k  \delta t, t +\delta t  \right) =  f_k \left(\bs{x},t\right) - \frac{1}{\tau_{LBM}} \bigg[f_k \left(\bs{x},t\right) - f_k^{eq} \left(\bs{x},t\right) \bigg]  + G_k \delta t,
    \label{eq:LBMEq_1}
\end{equation}
where $f_k^{eq}$ is the equilibrium distribution and the non-equilibrium distribution is given as $f_k^{non-eq} = f_k - f_k^{eq}$. The term $\bs{x}$ represents the position vector and $\bs{c}$ represents the discrete particle velocity, at time $t$. A schematic of the discrete $\bs{c}$ vectors is presented in Figure \ref{fig:lbmcvec} We employ a single relaxation time constant $\tau_{LBM}$ and $\delta t$ is the time increment in LBM. 
 
The distribution $f^{eq}_k$, in terms of the macroscopic fluid velocity $\tb{U}$ is expressed as: 
\begin{equation}
     f_k^{eq} = \rho \omega_k \bigg[ 1 + \frac{3 (\bs{c}_k \cdot \tb{U})}{c_s^2} + \frac{9 (\bs{c}_k \cdot \tb{U})^2}{2c_s^4} -  \frac{3 (\tb{U} \cdot \tb{U})}{2c_s^2}\bigg],
    \label{eq:f^eq}
\end{equation}
where $c_s$ represents the speed of sound in lattice units, $\rho$ is the fluid density, $\omega_k$ are the weights specific to the velocity set.

The discrete force term $G_k$ includes three forces. The dissipative drag force $G_D$, the buoyancy force $G_B$, and the shrinkage force $G_S$. The complete form is,
\begin{equation}
    G_k \left(\bs{x},t\right) =  \omega_k \bigg[ \frac{3 (\bs{c}_k -\tb{U})}{c_s^2} + \frac{9 (\bs{c}_k \cdot \tb{U})\bs{c}_k}{c_s^4} \bigg] \cdot \left(\tb{G}_D +\tb{G}_B +\tb{G}_S \right).
    \label{eq:discreteForce}
\end{equation}

The drag force $\tb{G}_D$ enforces the no-slip boundary condition at the solid-liquid interface and is elaborated as,
\begin{equation}
 \tb{G}_D\left(\bs{x},t\right) = - \frac{2\rho \nu h}{W_0} \left(1 -\phi_{liq}\right)^2 \tb{U},
\end{equation}
where $\nu$ denotes the kinematic viscosity, $h$ is a constant to adjust the physical units. Note that for the liquid phase, i.e., $\phi_{liq} = 1$, the  drag force will be zero. $W_0$ represents the thickness of the liquid-solid interface, and comes from Eq \ref{eq:Wpot}. 

The buoyancy force due to the change in the composition of the liquid phase is given by,
\begin{equation}
    \tb{G}_B\left(\bs{x},t\right) = -\rho \bs{g} \phi_{liq} \sum^{K-1}_{i=0}   \beta_i \left(C_i - C_i^{ref}\right),
    \label{eq:buoyancy}
\end{equation}
where $\bs{g}$ denotes the acceleration vector due to gravity, \{$\beta_i$\} are the solute expansion factors for each component, and $C_i^{ref}$ represents the a reference liquid concentration of the i$^{th}$ component. This formulation of force terms is similar to what is used by Takaki et al.\cite{Takaki2017} to incorporate natural convection in their model.


We have an additional term to incorporate shrinkage driven flow in our model. The shrinkage force arises from the difference in density between the solidified phase and the liquid phase. For a multiphase system as the interface is assumed to be a mixture of the phases, the shrinkage force for each phase is additive and has the form,
\begin{equation}
 \tb{G}_S\left(\bs{x},t\right) = \sum_{\alpha=0}^{N-1} \frac{(\rho_{\alpha}- \rho_{liq})}{\rho_{liq}} \frac{d\phi_\alpha}{dt} \bs{\hat{n}}_\alpha ;
 \hspace{1cm}
 \bs{\hat{n}}_\alpha = \frac{\nabla \phi_\alpha}{\Vert \nabla \phi_\alpha \Vert},
 \label{eq:shrinkage}
\end{equation}
where \{$\rho_{\alpha}$\} are the density of the phases and $\bs{\hat{n}}_\alpha$, is the normal vector at the interface. The term $(\rho_{\alpha}- \rho_{liq})/\rho_{liq}$ will also be referred to as $\eta_\rho$.

The macroscopic fluid density and velocity are then computed from the first and second moments of the distribution function as,
\begin{equation}
    \rho = \sum^{Q-1}_{k=0} f_k;
    \hspace{1cm}    
    \rho \tb{U} = \sum^{Q-1}_{k=0} f_k \bs{c}_k,
    \label{eq:rhovel}
\end{equation}

where $Q$ is the number of independent distribution functions. It is this macroscopic fluid velocity that is used in advecting the composition fields in Equation \ref{eq:conc}. Further details of the computational scheme are presented in the Appendix \ref{app:comScheme}.

\section{Results and discussion}

This section explores the simulation results and presents a discussion on the observations. First we investigate the solidification of a hypothetical binary system ($A-B$) in two distinct conditions, an isothermal condition where a single dendrite grows in a melt of constant undercooling (Case 1), and a Bridgman like directional solidification condition where we impose a constant thermal gradient $G$ and a constant pulling velocity $V$ (Case 2). Second, we investigate a hypothetical three-component ($A-B-C$) three-phase ($\alpha-\beta-\gamma$) eutectic system and its behavior under directional solidification (Case 3). The cases follow the simulation parameters mentioned in Appendix~\ref{app:simpara}, unless explicitly stated otherwise.

Note that we will refer to the gravity conditions directly by the buoyancy coefficients $\beta_i$ as a standard convention for multi-component systems. The scaling of buoyancy coefficients is equivalent to scaling the gravitational force by Eq. \ref{eq:buoyancy}. 
For the binary systems, (Case 1) and (Case 2), the convention is that the interdendritic liquid (lean in element $A$) is lighter than the far-field liquid of composition $C_0$, for positive values of $\beta_A$.

\subsection{Case 1 : Isothermal Solidification}
\label{subsec:case1}

\begin{figure}[ht]
    \centering
    \begin{subfigure}{.49\textwidth}
        \centering
        \includegraphics[width=\linewidth]{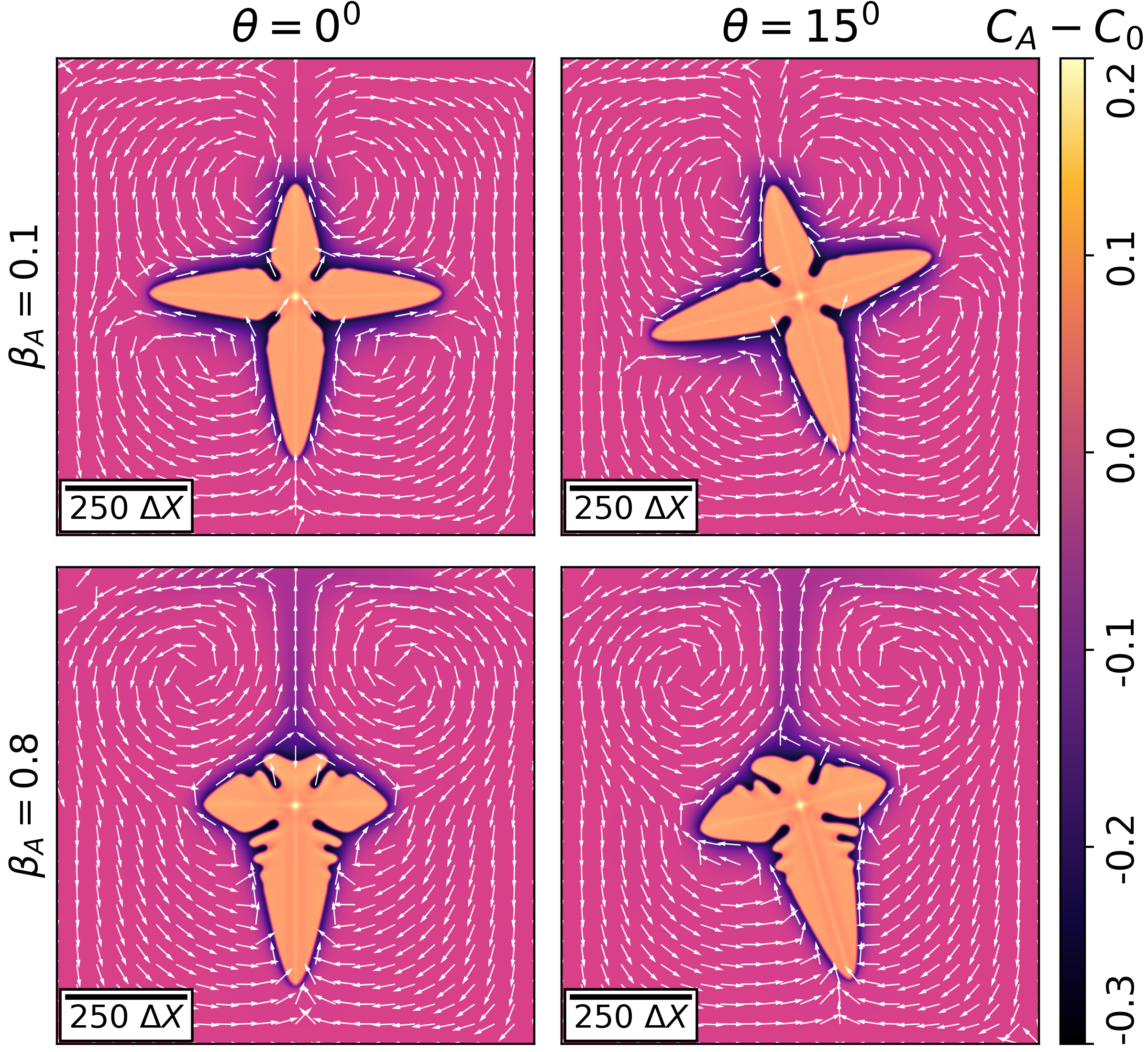}
        \caption{ }
        \label{fig:eqcimg}
    \end{subfigure}
    \hfill
    \begin{subfigure}{.49\textwidth}
        \centering
    \includegraphics[width=\linewidth]{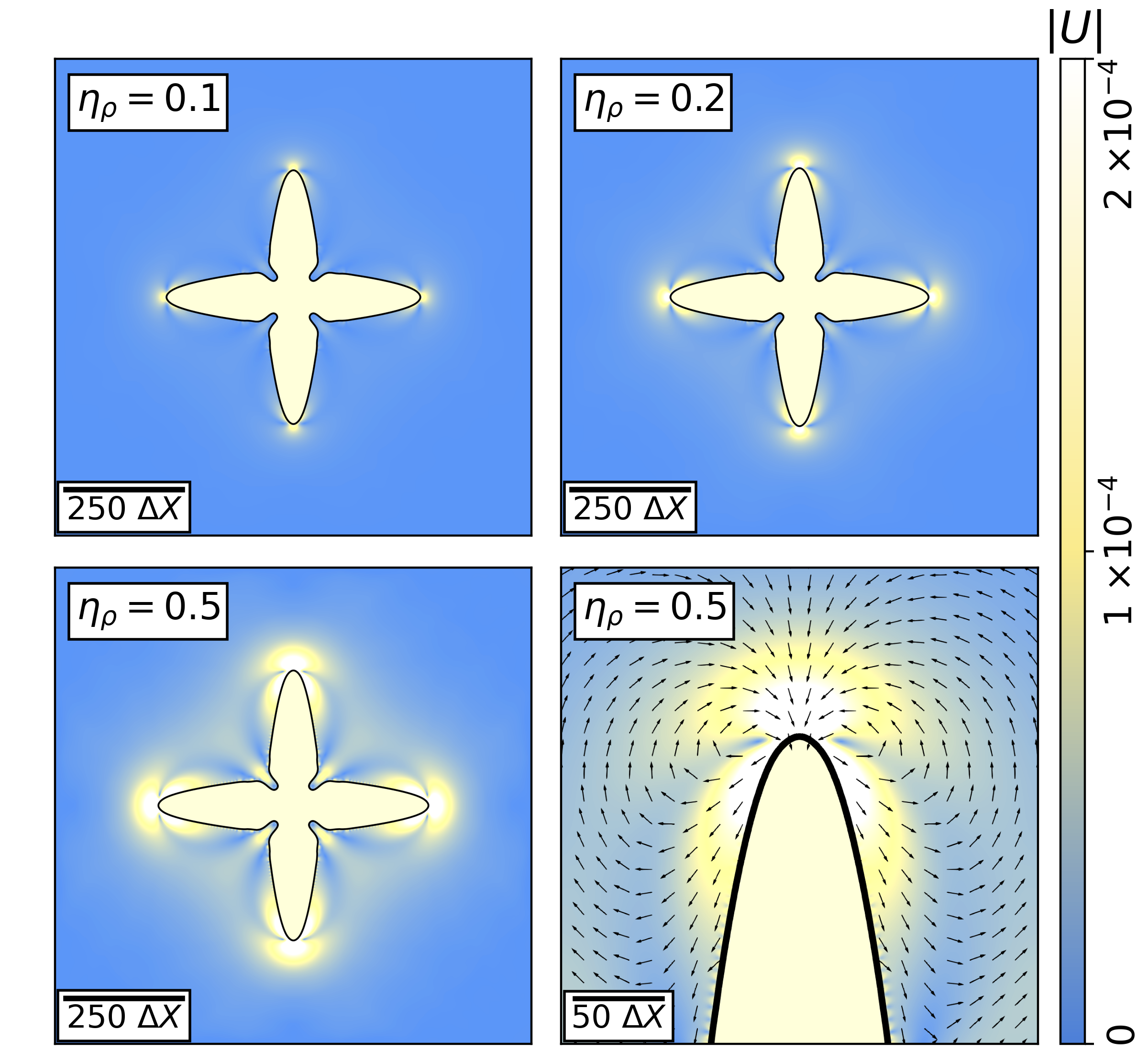}
        \caption{ }
        \label{fig:eqshrink}
    \end{subfigure}
    \caption{(a) Composition profiles of isothermally growing dendrites superimposed with normalized velocity vectors (white arrows). Here $\theta$ is the dendrite misorientation and $\beta_A$ is the buoyancy coefficient of the component $A$. 
    		(b) Velocity profiles around dendrites for different solid-liquid density ratios $\eta_\rho$. Here $\beta_A$ is set to $0$ to characterize only the effect of shrinkage.}
    \label{fig:iso_one}
\end{figure}

\begin{figure}    
     \centering
     \includegraphics[width=0.49\linewidth]{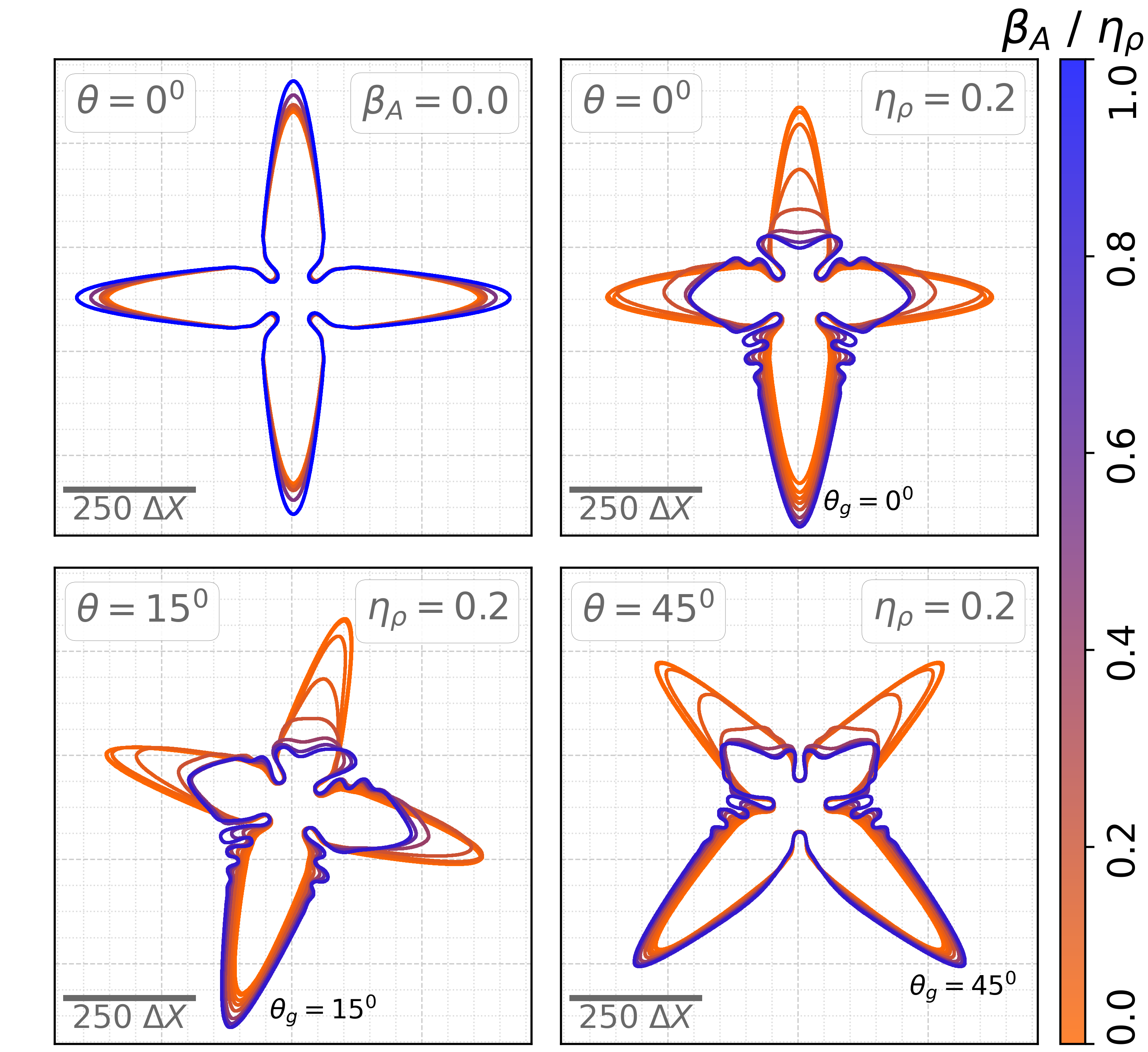}
     \caption{Superimposed contours of $\phi_{liq} = 0.5$ with varying $\beta_A$ and $\eta_\rho$. The top left subplot shows the effect $\eta_\rho$, i.e shrinkage driven flow on the dendrite contour. Others show the effect of $\beta_A$ for different dendrite misorientation angles $(\theta)$. The term $\theta_g$ is the orientation of an individual dendrite tip w.r.t gravity vector $\vec{g}$ (downwards).}
     \label{fig:eqcont}
\end{figure}

 \begin{figure}[ht]
     \centering
     \begin{subfigure}{.48\textwidth}
         \centering
         \includegraphics[width=\linewidth]{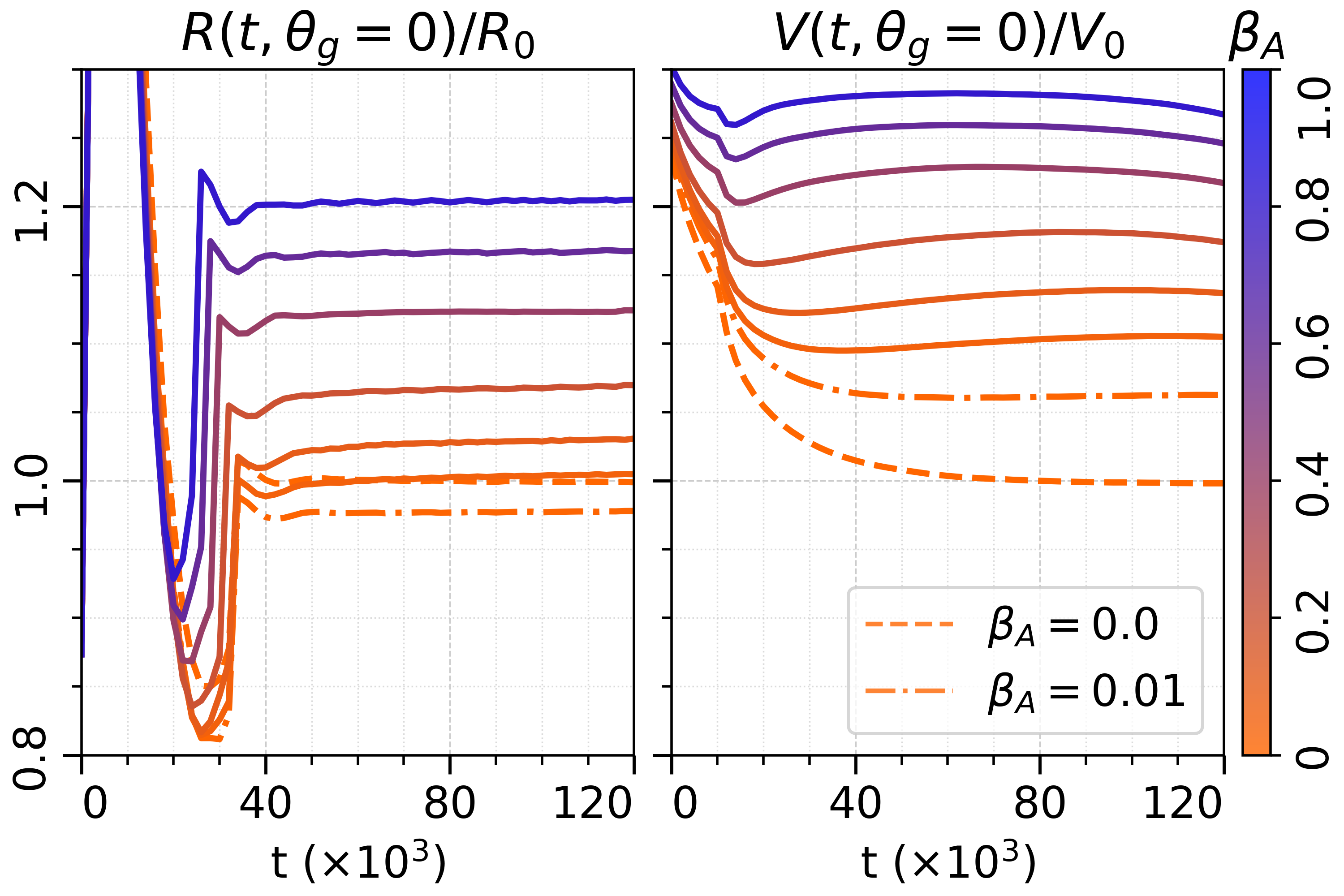}
         \caption{}
         \label{fig:eqvelradtime}
     \end{subfigure}
     \hfill
     \begin{subfigure}{.48\textwidth}
         \centering
     \includegraphics[width=\linewidth]{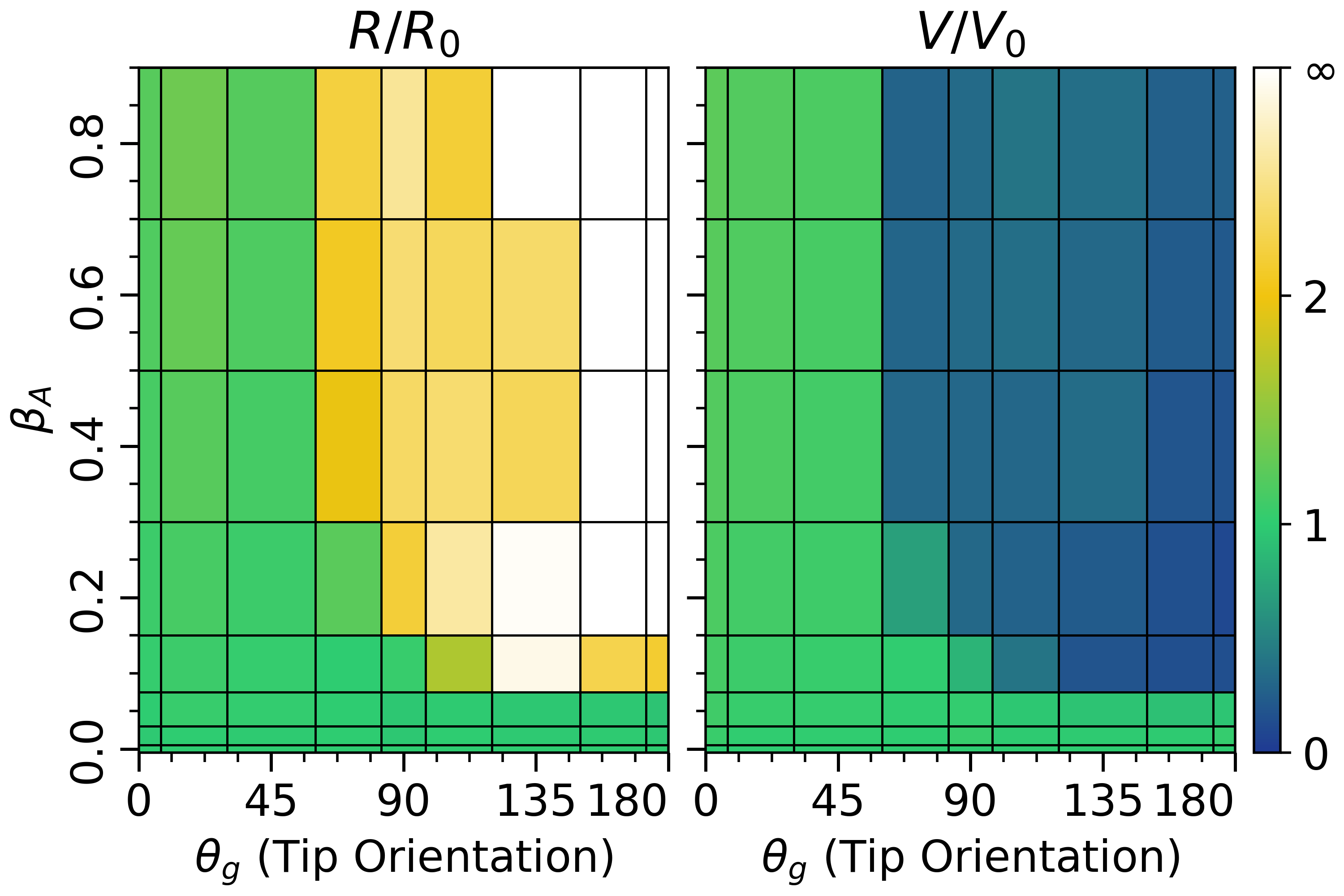}
         \caption{ }
         \label{fig:eqvelrad}
     \end{subfigure}
     \caption{  (a) Plots of scaled dendrite tip radius $R(t)$ and velocity $V(t)$ for the dendrite tip of orientation $\theta_g = 0$ (marked in the top right subplot of Figure \ref{fig:eqcont}).
                (b) A colored heatmap representation of the final steady state tip radius and velocity for different values of $\beta_A$ and $\theta_g$. The parameters of each dendrite arm in Figure \ref{fig:eqcont} is plotted here.
                Here $R_0$ and $V_0$ are the steady state values when $\beta_A = 0$, i.e. for no natural convection, and $\theta_g$ is the orientation of a single dendrite tip w.r.t $\vec{g}$.}
     \label{fig:iso_two}
 \end{figure}

The simulations of Case 1 are performed on a uniform grid of $1001\times1001$ cells, with a constant undercooling, and gravity acting downward. The phase diagram and other simulation parameters are listed in the Appendix \ref{app:simpara}. The boundary conditions for the fields $\phi$, $c$ and $\mu$ are Neumann, and for the LBM fields $f_i$ it is bounce back. 

The influence of natural convection on solidification of a single dendrite is shown in Figure \ref{fig:eqcimg}, which shows the composition profiles $C_A$ and normalized velocity vectors $|U|$ at a simulation time of $1.4\times10^4\Delta t$. We observe a strong asymmetric influence of convection on the shape of the growing dendrite. The dendrite arms pointed along $\vec{g}$, i.e. downward, experience a net flow towards the tips that enhances its growth by two mechanisms. First, flow brings in $A$ rich liquid from the far-field towards the growing tip, and second, it takes away the $A$-lean liquid from ahead of the tip to other regions above. Similarly, the arms pointed against $\vec{g}$, i.e. upward, experience an effective local decrease in $A$ as flow brings in $A$-lean liquid from other regions. These effects are stronger for the higher value of $\beta_A$ and so we see a stronger asymmetry of the dendrites for $\beta_A = 0.8$. In all cases the liquid ahead of the solidification front rises against $\vec{g}$ and forms plumes of $A$-lean liquid that drive further convection in the box. We also observe the preferential development of secondary arms as a response to incoming flow along certain favorably oriented arms. Similar observations on dendrite asymmetry and preferential growth were also made in the work of Takaki et.al\cite{Takaki2017}.

The influence of shrinkage driven convection is shown in Figure \ref{fig:eqshrink}. The plots reveal the magnitude of velocity field $|U|$ around the dendrites for different values of $\eta_\rho$. There we have kept $\beta_A = 0$ to neglect the effect of buoyancy. We observe that the regions of strong convection are located around the tips and the mean magnitude of flow increases with increasing $\eta_\rho$. The flow vectors ahead of the tip are directed towards it and so they bring in $A$ rich liquid from far-field, compensating for the shrinkage during growth.

The $\phi_{liq} = 0.5$ isosurfaces for various values of $\beta_A$, $\eta_\rho$ and $\theta$ are superimposed on each other and shown in Figure \ref{fig:eqcont}. The top left subplot shows the isosurfaces for $\beta_A = 0$. Here we see an increase in tip velocity caused by shrinkage driven flow. The extent of this increase is proportional to the density difference between the solid and the liquid phase ($\eta_\rho$). In the rest of the plots we see an effect of $\beta_A$ with varying $\theta$. We observe that the extent of asymmetry of the dendrite is proportional to the magnitude of $\beta_A$. The tips pointed against $\vec{g}$ are more sensitive to buoyancy driven convection, such that for larger values of $\beta_A$ those tips cease to grow and split.

To quantify this variation of the growth kinetics for individual tips, we define a orientation term $\theta_g$ to represent the angle between the tip and the gravity vector $\vec{g}$. For $\theta_g = 0$ (marked in Figure \ref{fig:eqcont}), the normalized profile of tip radius $R$ and velocity $V$ w.r.t time are shown in Figure \ref{fig:eqvelradtime}. The normalized steady-state tip radius and velocity for all the individual dendrite tips, are plotted as a heatmap in Figure \ref{fig:eqvelrad}, as a function of $\beta_A$ (Y-axis) and $\theta_g$ (X-axis). The point $\infty$ for radius, and the point $0$ for velocity marks the event of tip splitting. We observe in Figure \ref{fig:eqvelradtime} that for $\theta_g=0^\circ$ both $R$ and $V$ achieve a steady-state value and the values increase with $\beta_A$. For a given dendrite tip, the flux intensity factor $(\mathcal{F})$, which measures the solutal flux feeding the tip, follows the proportionality $\mathcal{F}^2 \propto RV^2$\cite{Isensee2022}. For dendrite tips favorably oriented w.r.t gravity, the the solutal flux towards the tip is much larger on account of the velocity vectors being directed towards the tip. In contrast, for the tips that are unfavorably oriented w.r.t gravity, i.e. at higher angles such as the horizontal or vertically growing tips, the velocity vectors are not all directed towards the tip. This fact either keeps the flux intensity the same or changes it to lower values, compared to the case without convection, depending on the value of $\beta_A$ and the inclination of the dendrite $\theta_g$. 

One can calculate the Peclet numbers $Pc = RV/2D$ (Non-dimensionalized) at the tips and we can see that for the tip directed downwards ($\theta_g=0^\circ$) the value is close 0.7 compared to the dendrite tip for $\theta_g=90^\circ$, where the value is 0.1 for $\beta_a=0.8$. For a substantial increase in the Peclet number, the microsolvability constant also changes substantially and this fact leads to both the radius and velocity to increase to larger values for the dendrite tip growing downwards. Conversely, for the tips growing at lower Peclet numbers the solvability constant is closer to the zero Peclet number limit, where the decrease in velocity is co-incident with an increase in the tip-radius. 





\subsection{Case 2 : Directional Solidification}
\label{subsec:case2}
Here we discuss the results of directional solidification from a planar interface. The simulations were performed on a computational grid of $1000 \times 5000$ cells, with an imposed upward thermal gradient and gravity acting downward. Other parameters are listed in Appendix \ref{app:simpara}. Boundary conditions on the left and right are periodic and for top and bottom are Neumann. We employ a moving frame approach to follow the dendritic growth front allowing us to simulate larger solidification distances. 

\begin{figure}[ht]
    \centering
    \begin{subfigure}{.48\textwidth}
        \centering
    \includegraphics[width=\linewidth]{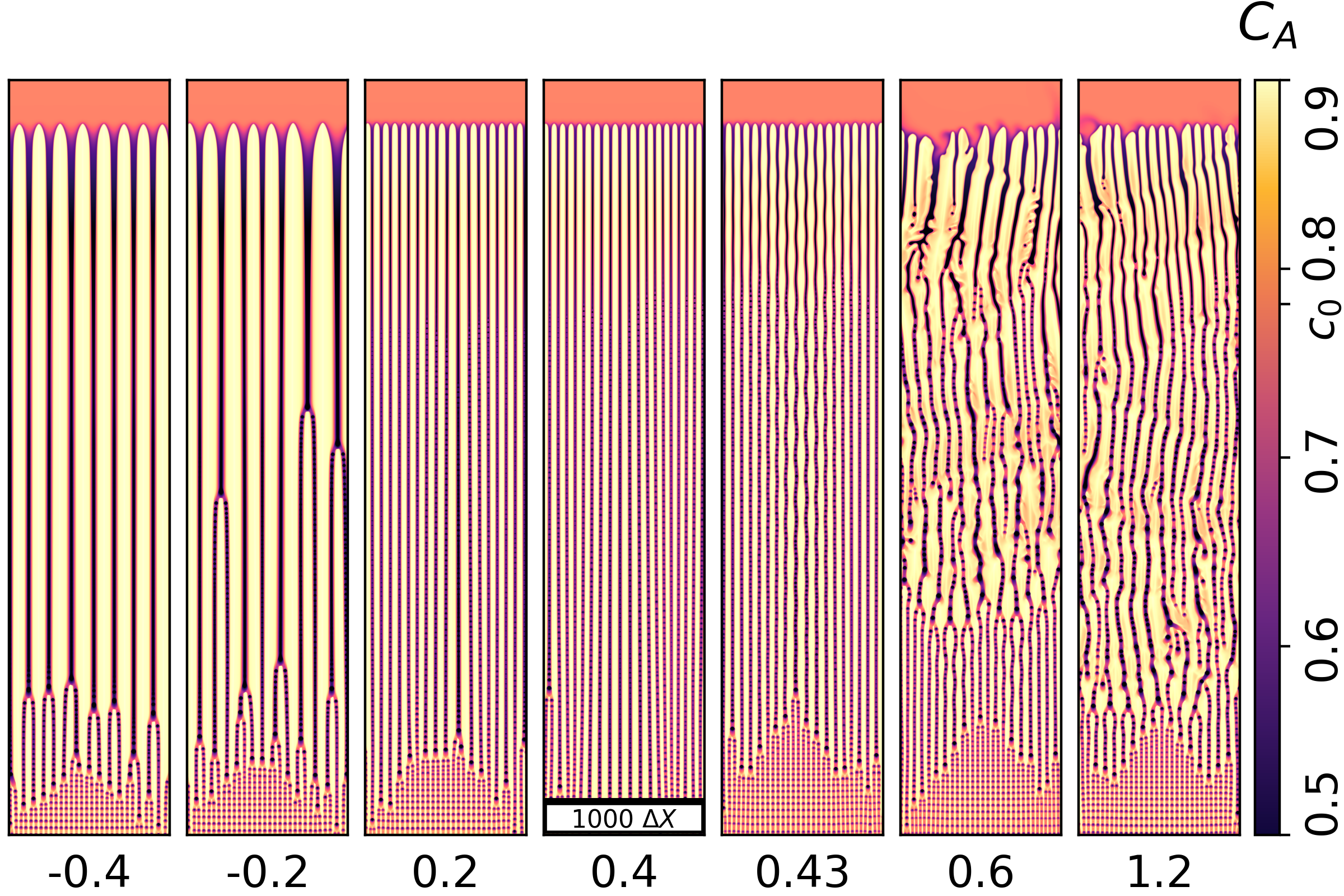}
        \caption{}
        \label{fig:dir_sims}
    \end{subfigure}
    \hfill
    \begin{subfigure}{.48\textwidth}
    \centering
    \includegraphics[width=\linewidth]{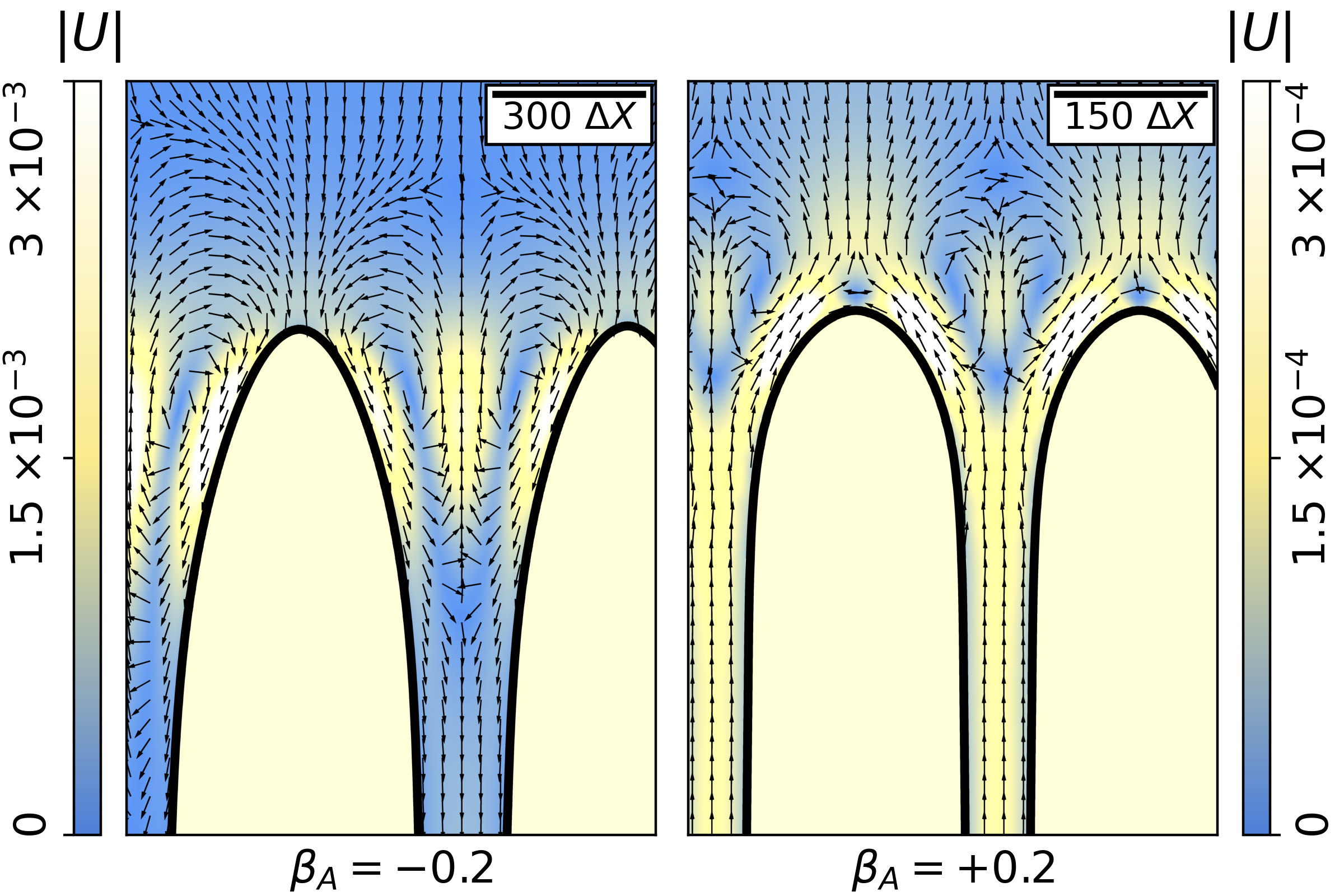}
    \caption{}
    \label{fig:dir_vels}
    \end{subfigure}
    \caption{(a) Composition profiles of directionally solidifying dendrites under different $\beta_A$ values marked in X-axis. (b) Velocity profiles around the dendrite tips (zoomed in) for two cases of (a).}
    \label{fig:dirsimvels}
\end{figure}

The entire composition profiles simulated microstructure, for certain values of $\beta_A$ are plotted in Figure \ref{fig:dir_sims}. We observe a significant influence of convection on the solidification morphology. For negative values of $\beta_A$, we see a larger dendritic arm spacing and tip radius, and for larger positive values $\beta_A \ge 0.6$ an emergence of instability in the growth front. The velocity fields corresponding to $\beta_A = +0.2$ and $\beta_A = -0.2$ are plotted in Figure \ref{fig:dir_vels}. We note that when $\beta_A$ is negative, the liquid ahead of the solidification front (lean in $A$) is heavier than the far field liquid and thus must sink. This creates a net downward flow just ahead of the dendrites, which feeds the far-field liquid to the tips. It also transports the $A$-lean liquid to the interdendritic region. In the case where $\beta_A$ is positive, the liquid ahead of the solidification front is lighter than the far-field liquid and must rise. This results in a net upward flow just ahead of the solid-liquid interface. In both cases there exist two stable vortices rotating in opposite directions between adjacent dendrites. The center of the vortices are at different locations for positive and negative values of $\beta_A$ w.r.t the dendrite tip. Additionally, the maximum flow velocity is higher for the case with negative $\beta_A$ (for the same magnitude). It can be attributed to the larger observed dendrite spacing and hence a lower friction experienced by flow. These results are identical to what was reported in\cite{Viardin2020,Takaki2020}.

The observed differences in the dendritic parameters are further investigated in the next sections.

\subsubsection{Effect on Kinetics}
\label{subsubsec:kinetics}
\begin{figure}[ht]
    \centering
    \begin{subfigure}{.48\textwidth}
        \centering
        \includegraphics[width=\linewidth]{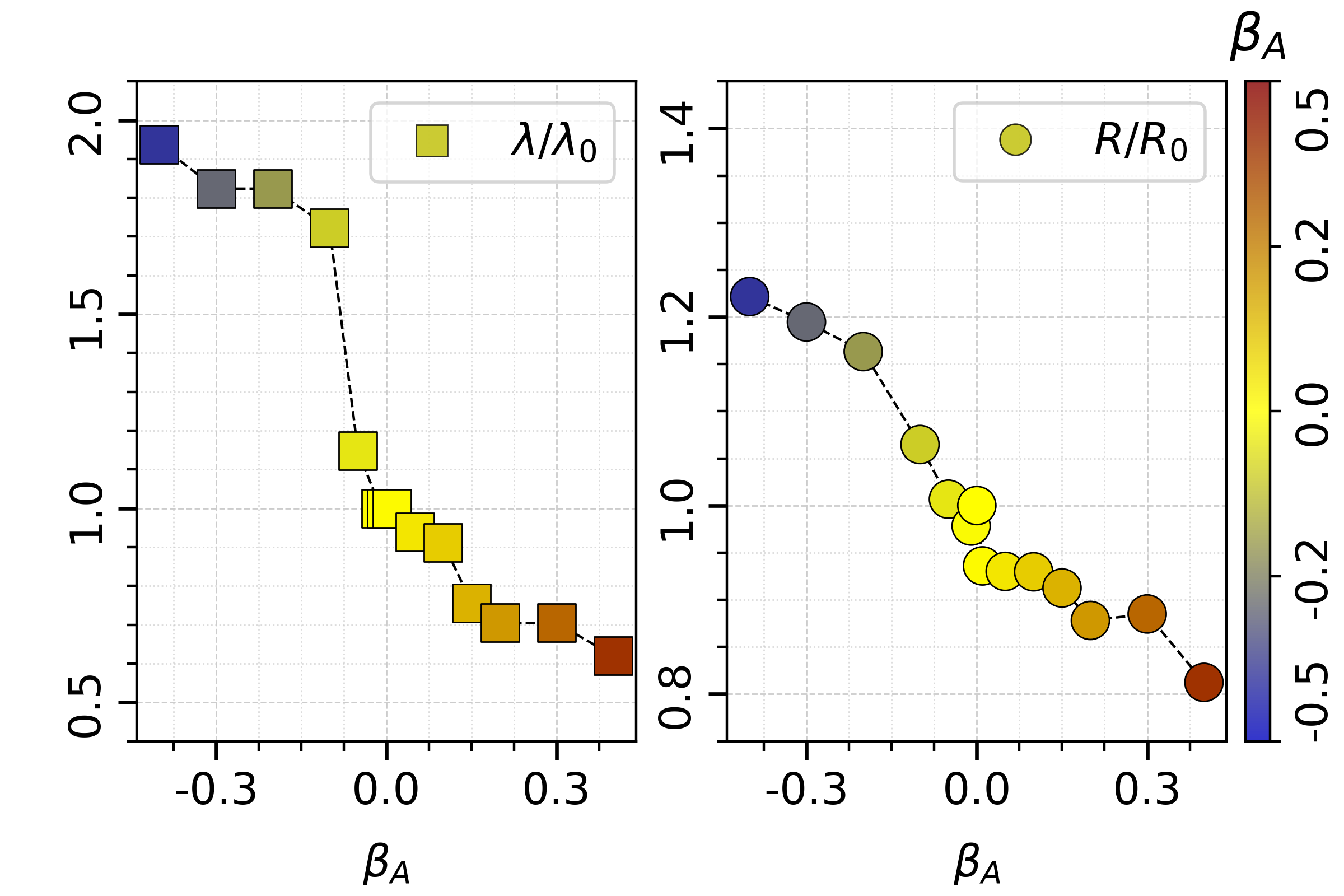}
        \caption{}
        \label{fig:dir_tradvel}
    \end{subfigure}
    \hfill
    \begin{subfigure}{.48\textwidth}
        \centering
    \includegraphics[width=\linewidth]{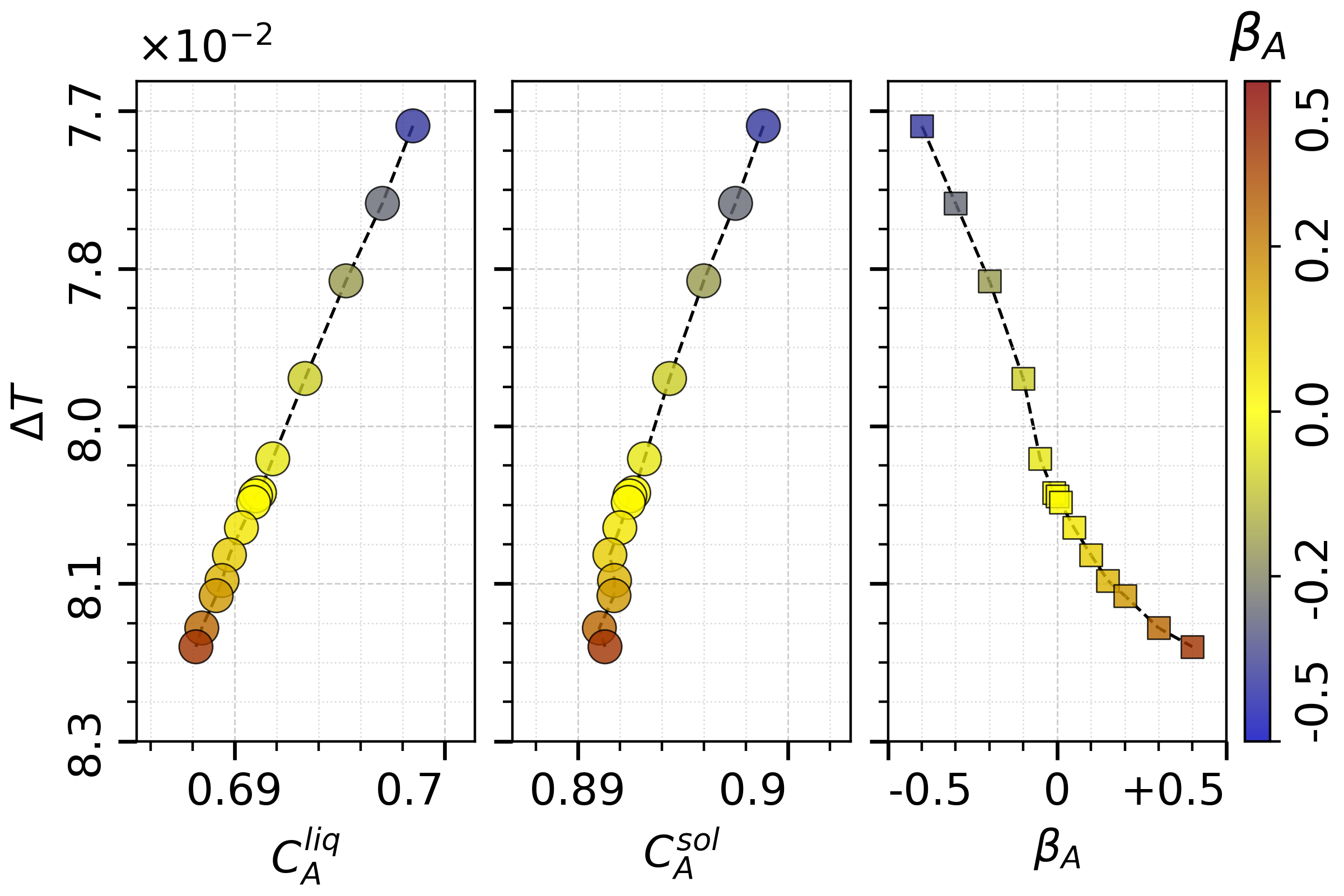}
        \caption{}
        \label{fig:beta_phasediag}
    \end{subfigure}
    \caption{(a) Primary-dendritic-arm-spacing ($\lambda/\lambda_0$) and tip radius ($R/R_0$) as a function of $\beta_A$. Here $\lambda_0$ and $R_0$ are the parameters for $\beta_A =0$. (b) Effect on $\Delta T$, $C_{liq}$ and $C_{sol}$.}
    \label{fig:betapdcomp}
\end{figure}



To quantify the variation of dendrite parameters in the stable growth regime, the simulations were performed on a larger grid of size $2500 \times 5000$ cells, with all the other conditions maintained same as mentioned. The primary-dendritic-arm-spacing PDAS $(\lambda)$ is measured as the mean distance between adjacent dendrite tips, and the tip radius $(R)$ is measured by fitting a parabola around the tips and calculating the radius of curvature at the vertex. The measured $\lambda$ and $R$ were normalized w.r.t to the values corresponding to $\beta_A = 0$, i.e. $\lambda_0$ and $R_0$, and the normalized values are plotted in Figure \ref{fig:dir_tradvel} as a function of $\beta_A$. We observe that as $\beta_A$ decreases (from positive to negative), the tip radius and PDAS increase. The extent of change in PDAS is larger for the negative values of $\beta_A$, than it is for positive values. This can be explained by the increase in $A$ lean liquid in the interdendritic region, due to convection and thus causing remelting in the interdendritic region and increasing the tip radius. 

Additionally there exists a variation of the tip undercooling $\Delta T$ w.r.t $\beta_A$. Convection alters the composition profiles around the dendrites, which causes the tips to grow at different undercoolings. This change of tip undercooling along with the solid and liquid compositions, $C_{sol}$ and $C_{liq}$ at the interface is plotted in Figure \ref{fig:beta_phasediag}. The compositions $C_{sol}$, $C_{liq}$ vary linearly with $\Delta T$ on account of the chosen phase diagram. We observe that for negative values of buoyancy coefficients both $C_{sol}$ and $C_{liq}$ increase and the tip grows at a lower $\Delta T$. This increase is caused by the velocity vectors being pointed toward the tip. Similarly the flow vectors being directed away from the tip, in the $\beta_A > 0$ regime causes the values of $C_{sol}$, $C_{liq}$ to decrease and the tips grow at larger undercoolings. This jump in undercooling is higher for negative $\beta_A$ due to the mean flow velocity being higher in this regime. 


\subsubsection{Emergence of Instability}
\label{subsubsec:stability}

\begin{figure}[ht]
    \centering
    \begin{subfigure}{.48\textwidth}
        \centering
        \includegraphics[width=\linewidth]{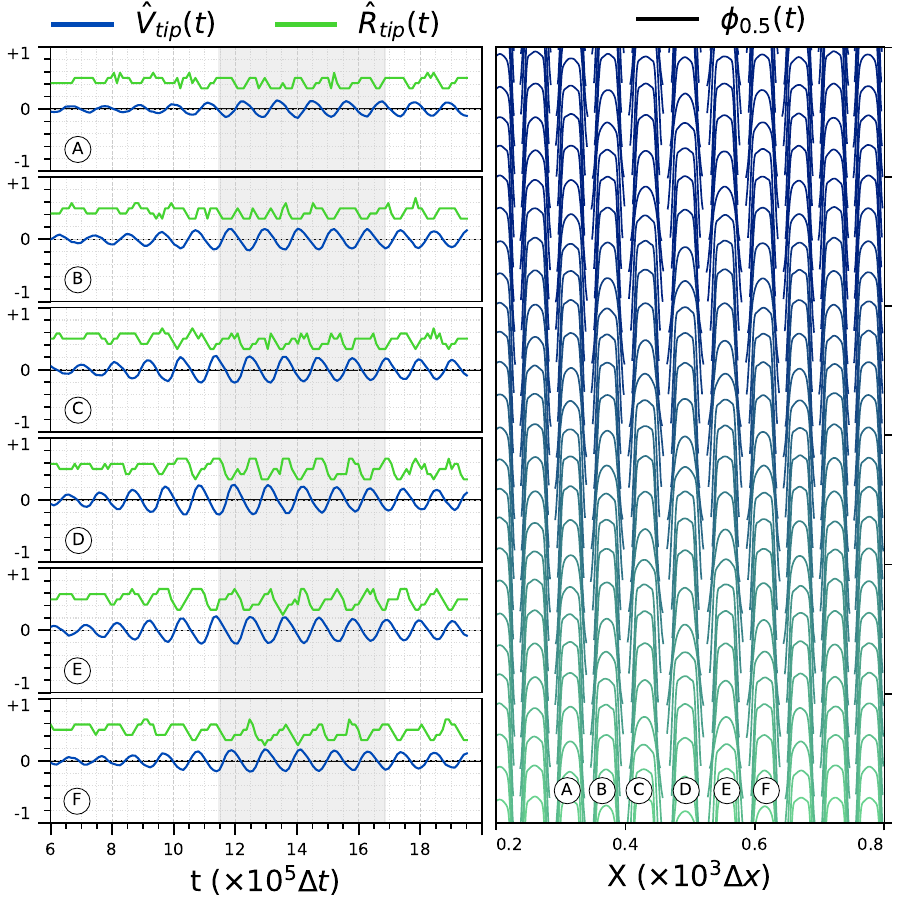}
        \caption{$\beta_A = 0.43$}
        \label{fig:RV_beta_43}
    \end{subfigure}
    \hfill
    \begin{subfigure}{.48\textwidth}
        \centering
    \includegraphics[width=\linewidth]{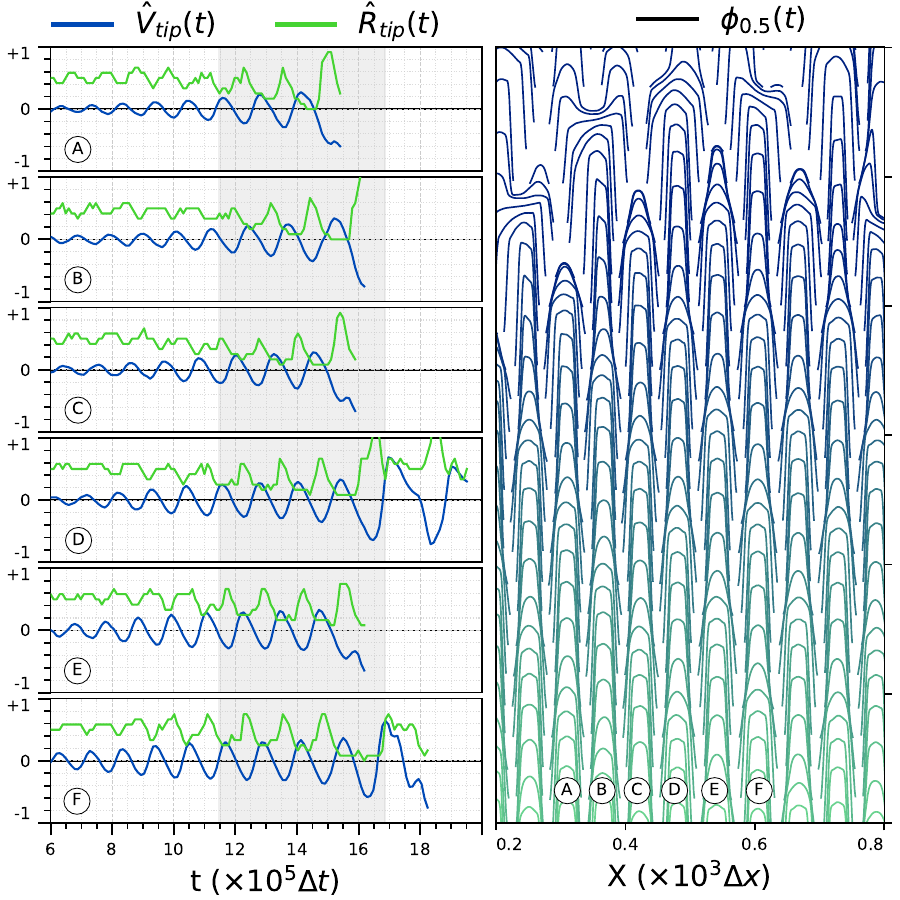}
        \caption{$\beta_A = 0.44$}
        \label{fig:RV_beta_44}
    \end{subfigure}
    \caption{Amplitude-bounded (a) and unstable (b) oscillations in the growth of columnar dendrites. Here, $\hat{V}_{tip} = V_{tip}-V_{pull}$ is the tip velocity in the moving frame and $\hat{R}_{tip} = R_{tip}/R_{ref}$ is scaled tip radius.}
    \label{fig:RV_beta}
\end{figure}

Beyond a critical $\beta_A$, the dendritic growth front destabilizes and shows tip oscillations, as is seen for $\beta_A \ge 0.6$ in Figure \ref{fig:dir_sims}. Similar kind of tip oscillations for the Ti-Al system, at a lower Peclet number regime was studied in the work of Viardin et. el.\cite{Viardin2020}. Oscillations were also observed in directional growth of CMSX-4 both computationally\cite{Yang2019} and experimentally\cite{Reinhart2020}. 

We, in this section investigate the emergence of this instability as controlled by the buoyancy coefficient $\beta_A$. We focus on a small range of $\beta_A$ values in which the solidification front exhibits stable oscillatory behavior. This is presented in Figure \ref{fig:RV_beta} for $\beta_A$ values of $0.43$ and $0.44$. For each figure, the box on the RHS shows the dendritic contours ($\phi_{liq}=0.5$) at different time intervals, spatially superimposed on each other. Only a certain contour region around the tips is plotted for clarity. As the front moves with a constant imposed velocity, each of the contours advance in Y-axis (vertical) and their oscillatory behavior is observed in the periodic change of their shape. Correspondingly, a few dendrite tips are identified and labeled with the alphabets $A-F$, and their radii and velocities are tracked and plotted as a function of time on the LHS plots. Note that the time interval corresponding to the Y-axis range of the RHS plots is marked with a gray background in the LHS figure. Tip velocity $V_{tip}$ is represented in the frame moving with $V_{pull}$ as $\hat{V}_{tip} = V_{tip}-V_{pull}$, and $R_{tip}$ is scaled as $\hat{R}_{tip} = R_{tip}/R_{ref}$ for clarity. 

Both the tip radius and the tip velocity show oscillatory behavior with an equal wavelength, frequency, and a constant phase difference of $\pi$. An increase in tip velocity is accompanied by a decrease in tip radius and vice versa. The amplitude of these oscillations remains bounded for $\beta_A = 0.43$. At and beyond $\beta_A = 0.44$, the oscillation amplitude increases until it reaches the point of instability where certain tips stop growing, and others undergo splitting events. There is a spatial coherence to these oscillations. Adjacent dendrites show oscillations at a constant $\pi$ phase difference; this coupling arises from the interaction of the vortices associated with the tips, which is discussed next. This coherence which causes alternate (2nd neighbors) dendrites to oscillate in phase w.r.t each other, also causes the death of alternate dendrite tips in coherence. This is observed in the RHS plot of Figure \ref{fig:RV_beta_44}, where alternate tips, A,C,E, etc. die almost simultaneously and B,D,F survives, additionally the surviving tips undergo splitting events later. 

\begin{figure}[!htbp]
    \centering
    \begin{subfigure}{.48\textwidth}
        \centering
        \includegraphics[width=\linewidth]{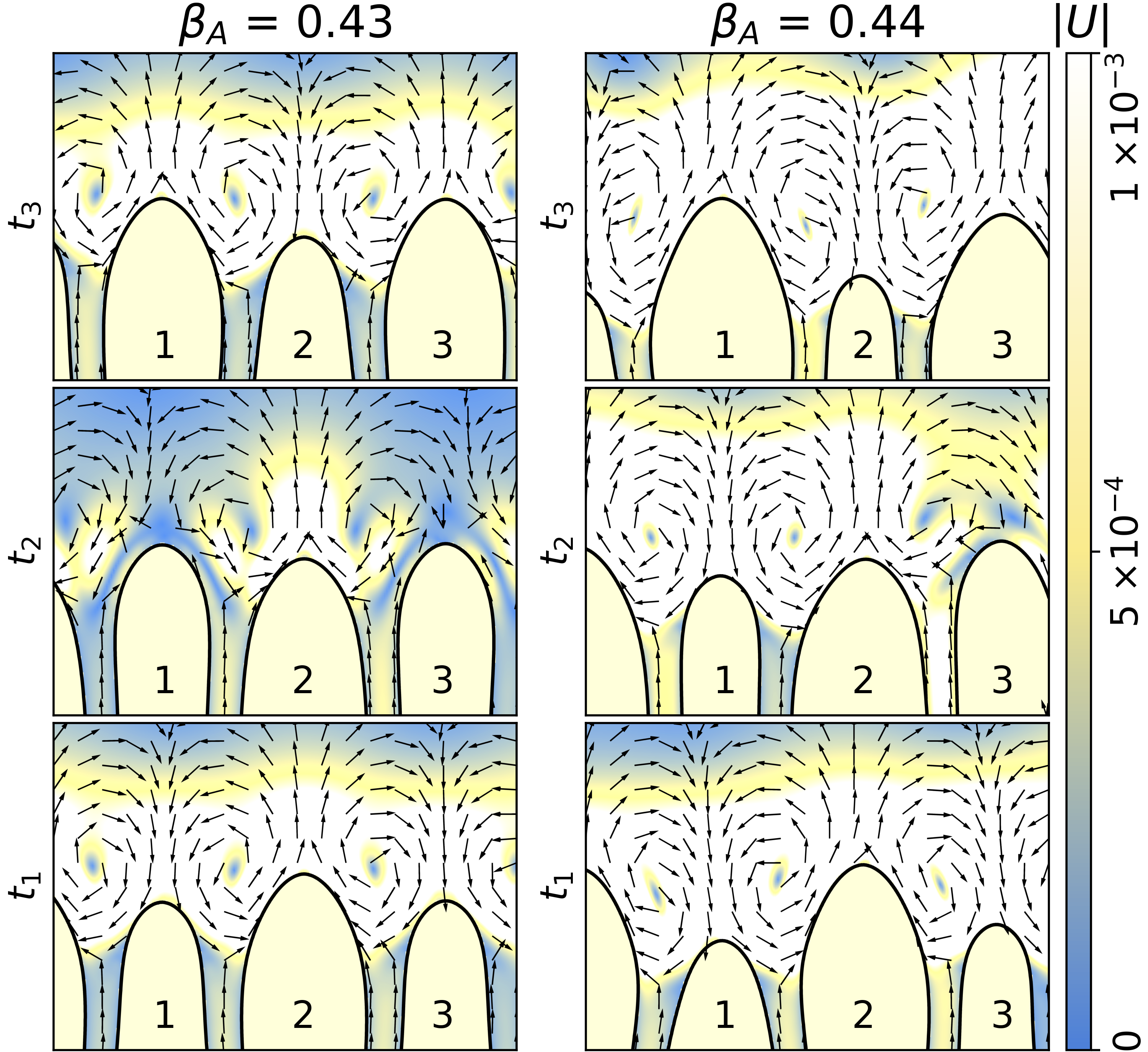}
        \caption{}
        \label{fig:dir_vec_oscii}
    \end{subfigure}
    \hfill
    \begin{subfigure}{.48\textwidth}
        \centering
    \includegraphics[width=\linewidth]{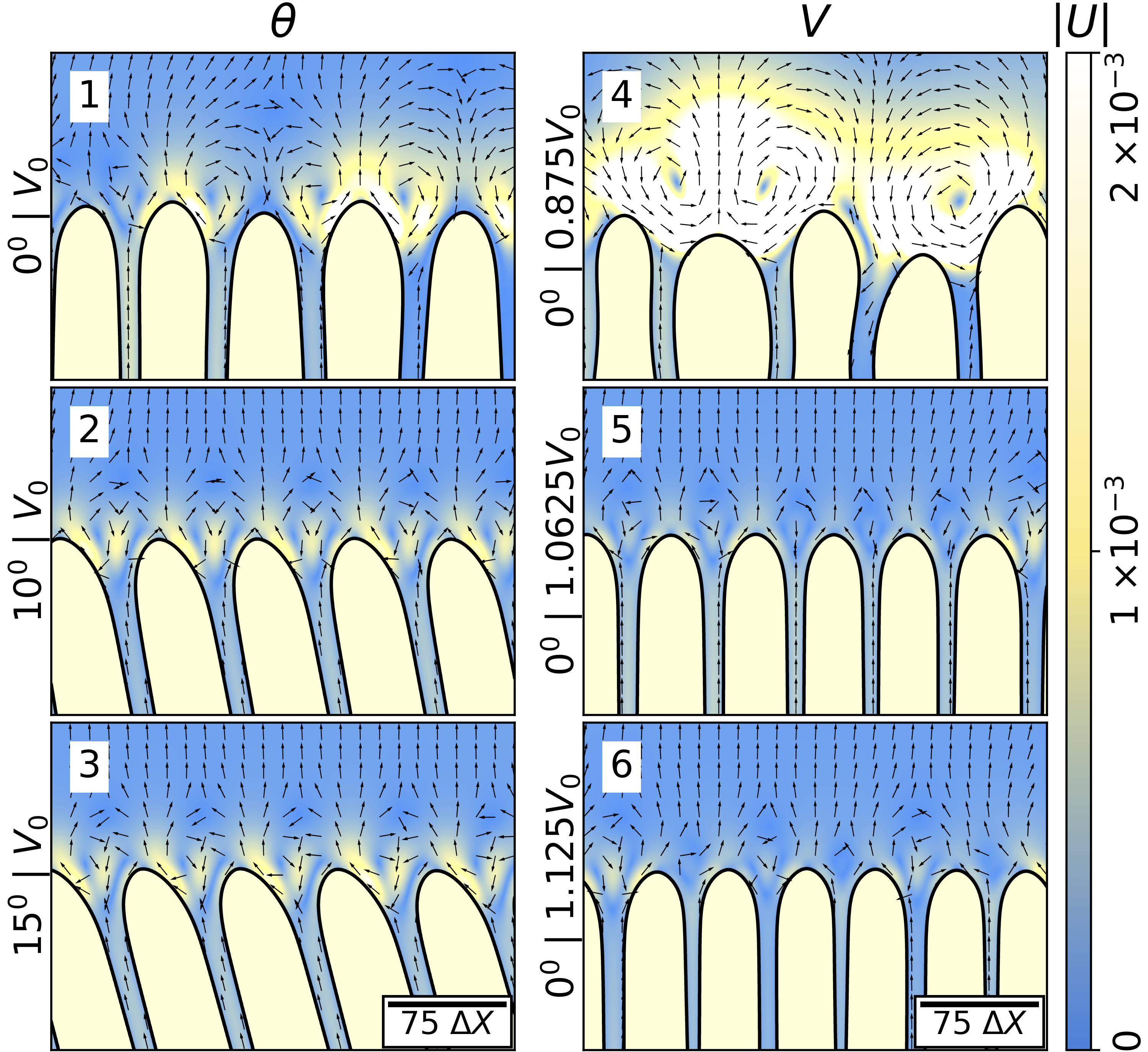}
        \caption{}
        \label{fig:dir_stability}
    \end{subfigure}
    \caption{Velocity field around dendrites. (a) Shows flow dendrite interaction during tip oscillations. The times $t_1 < t_2 < t_3$ marks consecutive instances on a oscillation half period. (b) Shows the stabilizing effects of tilt and pulling velocity for $\beta_A = 0.5$ at $t = 5.4\times10^5$ (early stage before the development of instabilities).}
    \label{fig:beta_vec_stab}
\end{figure}

The Figure \ref{fig:dir_vec_oscii} shows the velocity fields around dendrites at different points during the oscillation cycle for $\beta_A$ values of $0.43$ and $0.44$. The times $t_1 < t_2 < t_3$ marks consecutive instances on a oscillation half period. The dendrites marked 1 and 3 at time $t_1$, are in the same point of the oscillation cycle, as the dendrite marked 2 at time $t_3$. For dendrites marked 1 and 3, $t_1$ is at the point of minimum tip radius and maximum velocity and $t_3$ is at the opposite conditions. $t_2$ shows an intermediate state. Our first observation is that, unlike the steady-state conditions (in Figure \ref{fig:dir_vels}), the vortices associated with dendrite oscillations have a size comparable to the local PDAS i.e there exists one vortex between a pair of adjacent dendrites. This immediately creates alternate flow patterns ahead of adjacent tips and this couples the tip growth kinetics to the spatial distribution of the vortices. Therefore, the spatial coherence of tip oscillations, discussed earlier, were not observed when the vortex size was on average larger than the local PDAS, in a low Peclet number solidification regime\cite{Viardin2020, Yang2019}.

A state minimum tip radius is associated with the flow being directed toward the tips. We gather from the section \ref{subsubsec:kinetics} that when flow is directed towards the tips, which is a situation associated with negative buoyancy coefficients, the dendrites achieve a state of higher tip radius and lower tip undercooling. And so at time $t_1$, when the local flow conditions around the dendrite tips 1 and 3, mimic a $\beta_A < 0$ condition, the tips grow to a state of larger tip radius and lower undercooling. As the tips grow and surpasses their immediate neighbors, the vortices flip their rotation and at a time $t_3$, the tips experience a velocity field pointed away from it. This creates local conditions similar to that of $\beta_A > 0$ around the tips and they grow to achieve a state of low tip radius and higher tip undercooling. The stability of this dynamic relationship is dependent on the magnitude of $\beta_A$. A higher value of $\beta_A$ (higher buoyancy forces) leads to a larger amplitude of the oscillations in tip radii and velocity, and beyond a threshold the steady coupling between the tip kinetics and the vortices is lost. The pattern enters an unstable regime of large amplitude oscillations accompanied by frequent events of tip splitting. This is observed at later part of Figure \ref{fig:RV_beta_44}, after a simulation time of $t > 14\times 10^5$, and the top section of the contour plots. The phenomena is also seen in the composition plots for $\beta_A = 0.6$ and $1.2$ in Figure \ref{fig:dir_sims}.

\subsubsection{Effect of Control Parameters}
\label{subsubsec:controls}

\begin{figure}[ht]
    \centering
    \begin{subfigure}{.48\textwidth}
        \centering
        \includegraphics[width=\linewidth]{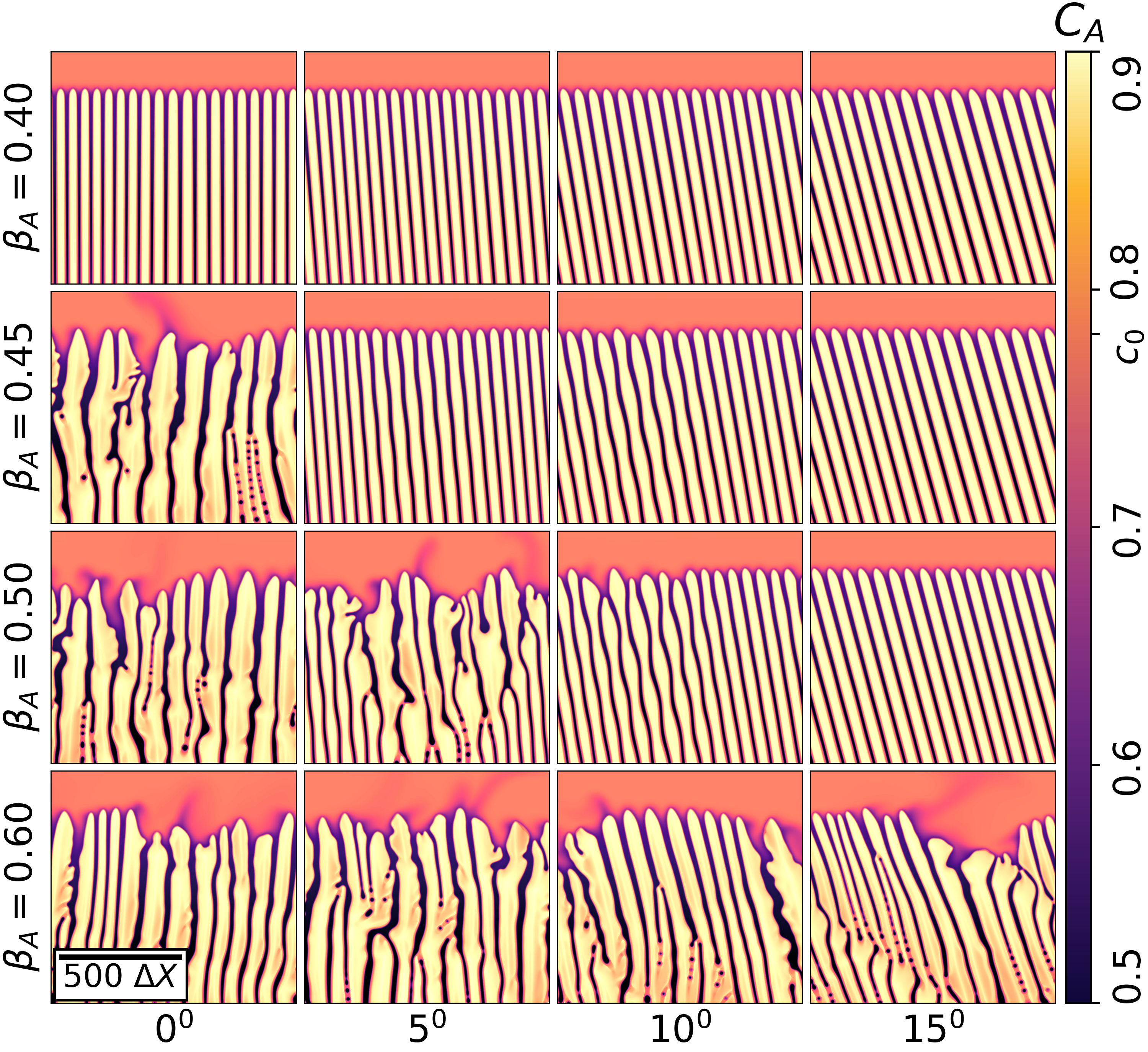}
        \caption{Effect of seed tilt $\theta$.}
        \label{fig:beta_tilt}
    \end{subfigure}
    \hfill
    \begin{subfigure}{.48\textwidth}
        \centering
    \includegraphics[width=\linewidth]{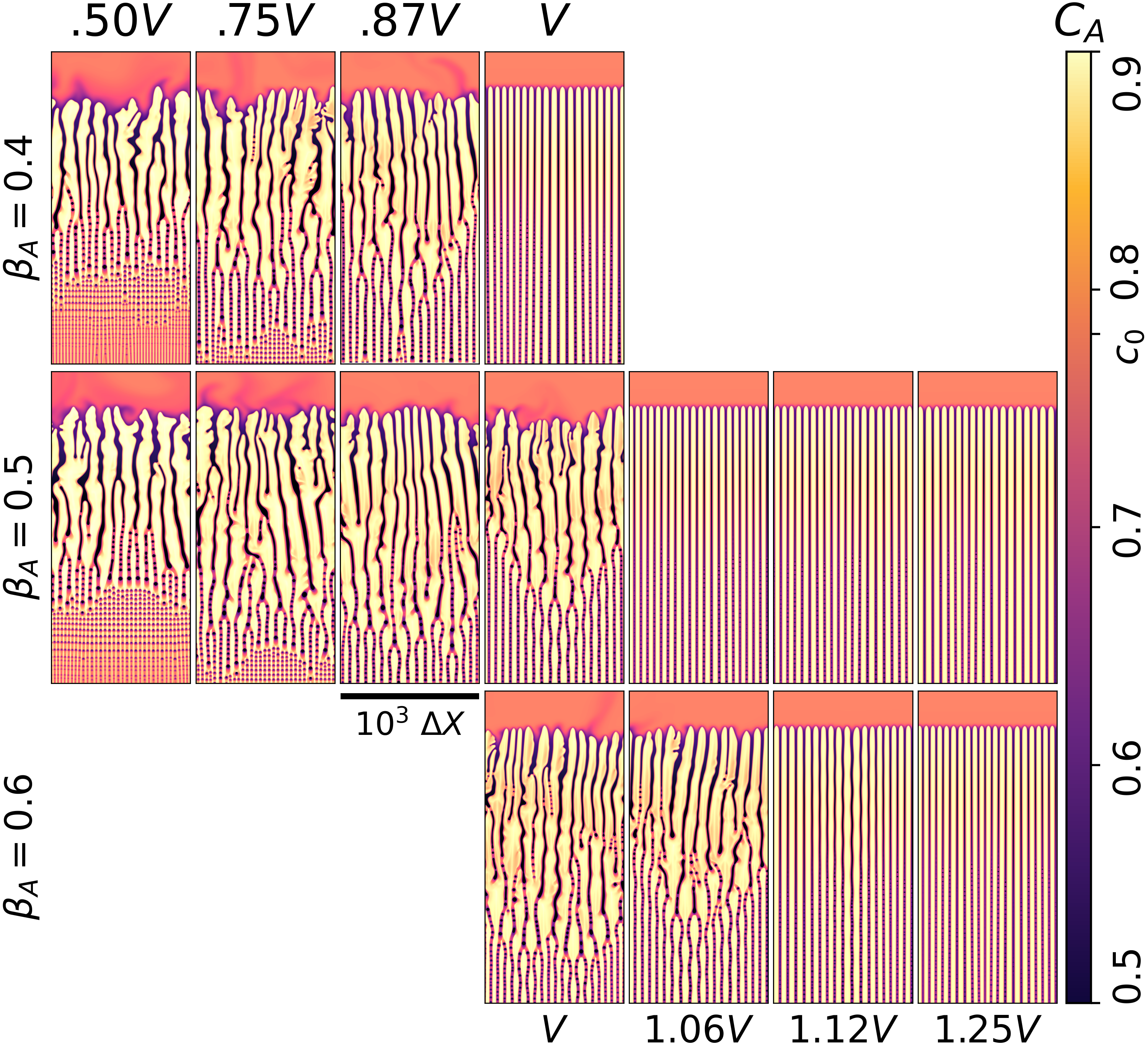}
        \caption{Effect of pulling velocity $V$.}
        \label{fig:beta_vpull}
    \end{subfigure}
    \caption{Snapshots of composition fields for varying values of $\beta_A$ (Y-axis) and (a) seed tilt $\theta$ (X-axis) and $t = 1.2\times10^6\Delta t$, and (b) pulling velocity $V$ (X-axis) and $t = 1.4\times10^6\Delta t$.}
    \label{fig:dir_controlparams}
\end{figure}

In Figure \ref{fig:beta_tilt} we investigate the flow patterns associated with dendrites growing at a tilt $\theta$ w.r.t. the thermal gradient. The plots show snapshots of the composition profiles (zoomed in) for simulations at different values of $\beta_A$ and $\theta$. For $\beta_A \le 0.4$, the solidification front does not destabilize and for $\beta_A \ge 0.6$ it destabilizes regardless of the seed tilt $\theta$. For the bounded range of values, $\beta_A \in (0.4, 0.6)$, we observe that an increase in seed tilt stabilizes the solidification front. This is explained by a net decrease in the vertical component of $\textbf{U}$ around the tips, which reduces the effective force acting on the solute boundary layer and prevents the front front from destabilizing. This change of velocity profile is shown in Figure \ref{fig:dir_stability} column $\theta$ (left). We observe that tilting the dendrite growth direction changes the velocity profiles in such a way that each dendrite tip experiences a similar flow pattern ahead of it, a condition akin to steady-state growth of Figure \ref{fig:dir_vels}. Note that is stabilization effect was not observed in the simulations of Yang et. al.\cite{Yang2019} where tip branching mechanisms were prominent.

In Figure \ref{fig:beta_vpull} we investigate the flow patterns associated with dendrites growing at different pulling velocities $V$. The stabilizing effect of higher pulling velocities (or cooling rates) has been experimentally observed and is well known to researchers, and it is reproduced in our simulations. An increase in pulling velocity is found to stabilize the growth front by two mechanisms. First, reducing the PDAS which reduces permeability and, in turn, the mean velocity in the region. Second, by reducing the diffusion length (because of an increase in velocity), there is a decrease in the thickness of the solute boundary layer, and thereby the driving force for buoyancy-driven flow diminishes. This can be observed in the velocity fields around the dendrite tips in the Figure \ref{fig:dir_stability} column $V$ (right).

\subsection{Case 3 : Eutectic Solidification}
\label{subsec:case3}

In this section, we investigate the convection effects on the directional solidification behavior of a hypothetical three-phase three-component eutectic system with a symmetric phase diagram and equal diagonal inter-diffusivity values, details of which are mentioned in the Appendix \ref{app:simpara}. For the discussion, we limit ourselves to the three simplest regular patterns, $\alpha\beta$, $\alpha\beta\gamma$, and $\alpha\beta\alpha\gamma$. The influence of flow is investigated by examining changes in the undercooling $(\Delta T)$ vs. spacing $(\lambda)$ relationships as a function of the elemental buoyancy coefficients $\beta_A$ and $ \beta_B$ (collectively referred to as $\beta_{A,B}$), where spacing $\lambda$ is the length of one period. In addition, the differences in the oscillatory modes are highlighted. Note that the pattern images and curves will be directly referred to by their $\beta_{A,B}$ values from here on, within the unique context of the respective Figure.

\begin{figure}[ht]
    \centering
    \begin{subfigure}{.49\textwidth}
        \centering
        \includegraphics[width=\linewidth]{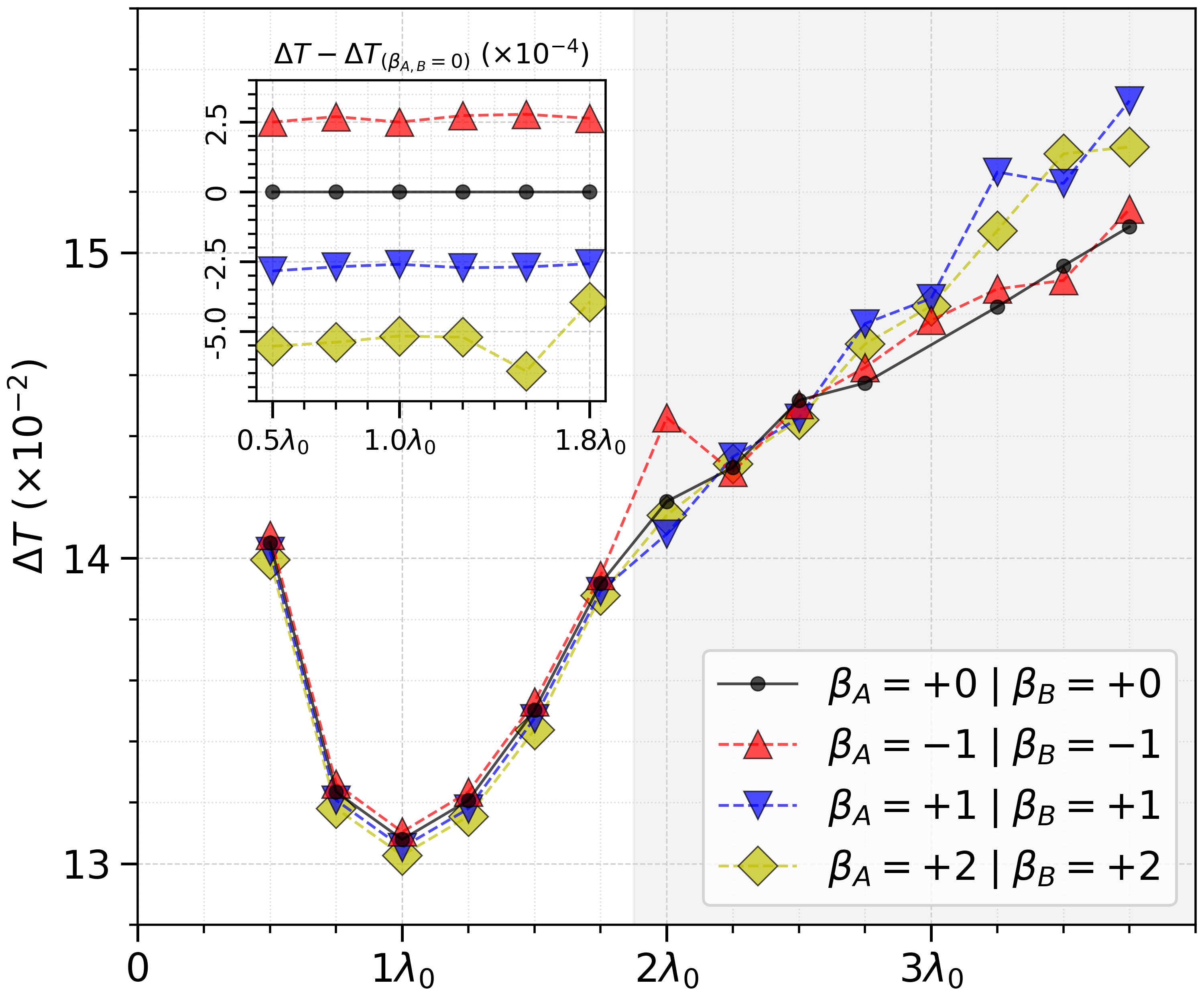}
        \caption{$\alpha\beta$}
        \label{fig:eu2_LdT}
    \end{subfigure}
    \hfill
    \begin{subfigure}{.49\textwidth}
        \centering
    \includegraphics[width=\linewidth]{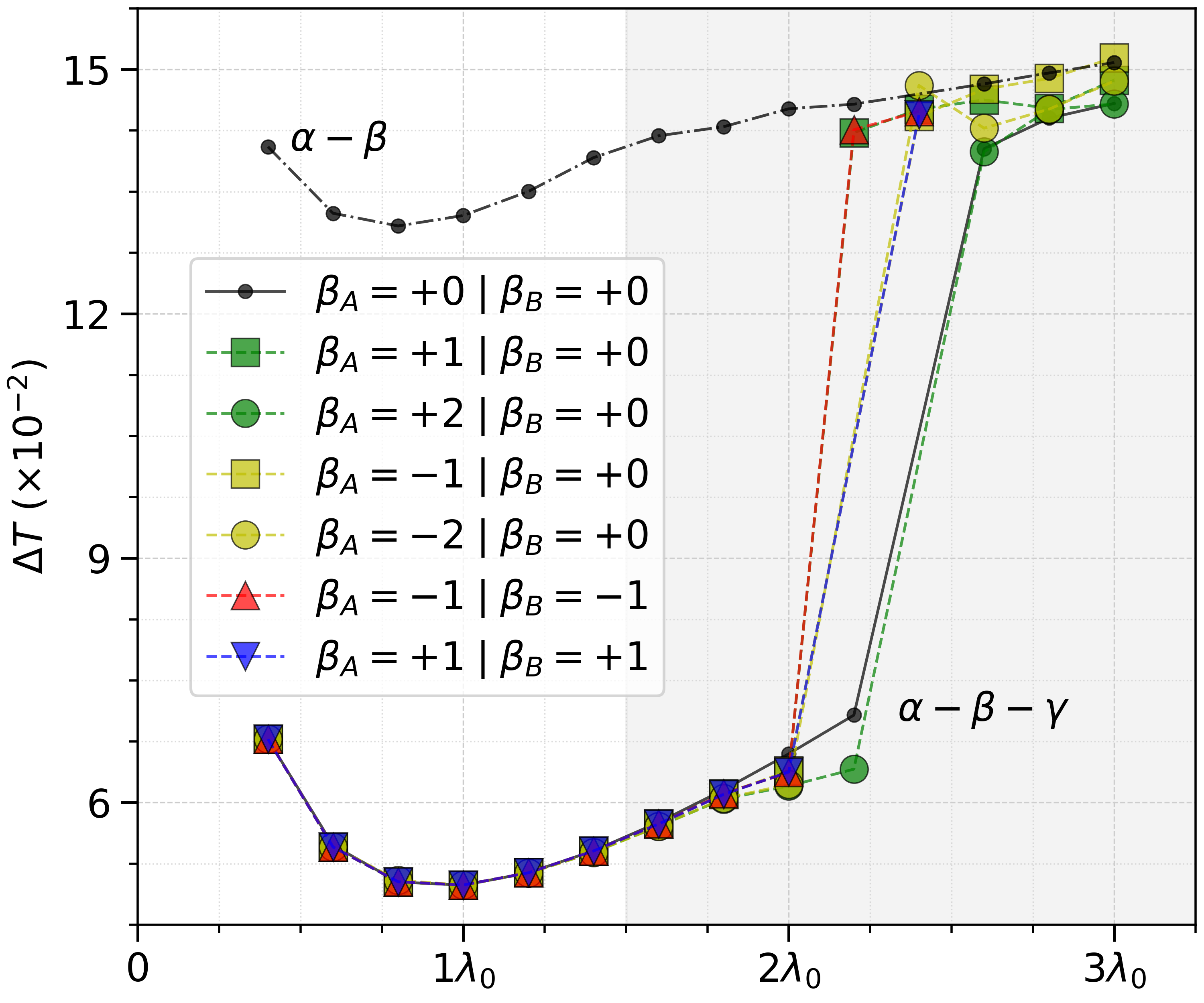}
        \caption{$\alpha\beta\gamma$}
        \label{fig:eu3_LdT}
    \end{subfigure}
    \caption{Steady state $\lambda$ vs $\Delta T$ plots for (a) binary eutectic system (regular $\alpha\beta$) and (b) ternary eutectic system (regular $\alpha\beta\gamma$), under different $\beta_{A,B}$ influence. The grey background marks the region where the solidification front shows oscillatory behavior.}
    \label{fig:euLdT}
\end{figure}

To compute the $\lambda$ v/s $\Delta T$ relationships, we have performed simulations on a grid of size $(4\lambda \times 700)$ cells, with a constant thermal gradient, a constant pulling velocity, and gravity acting downwards. The boundary conditions on the left and right are periodic, and those on the top and bottom are Neumann. We have chosen $\lambda_0$ to be the length-scale corresponding to that of minimum undercooling for the respective pattern. 

\subsubsection{Undercooling v/s spacing relationships}
\label {subsubsec:eu-LdT}

Figure \ref{fig:eu2_LdT} plots the $\lambda$ v/s $\Delta T$ curves for the regular two-phase $\alpha\beta$ pattern for different values of $\beta_{A,B}$. It is observed from the simulations that beyond a spacing of $\sim 1.9\lambda_0$ the $\alpha\beta$ pattern enters an oscillatory regime. Below this threshold, there exists a simple relationship between $\beta_{A,B}$ and the attained steady-state $\Delta T$, shown in the subplot of Figure \ref{fig:eu2_LdT}. Negative buoyancy coefficients of $[-1,-1]$ causes an increase in $\Delta T$, whereas positive values of $[1,1]$ and $[2,2]$ causes a decrease. The strength of this influence is proportional to the magnitude of $\beta_{A,B}$.  In this regime, the undercoolings follow a simple ordered relationship of $\Delta T_{[-1,-1]} >  \Delta T_{[0,0]} >  \Delta T_{[1,1]} >  \Delta T_{[2,2]}$. Note that the actual magnitude of the differences is of the order $10^{-4}$ (Non-dimensionalized temperature units). 

This relationship of $\Delta T$ is not observed when the $\alpha\beta$ pattern enters the oscillatory regime $(\lambda > 1.9\lambda_0)$. Both the curves corresponding to positive buoyancy coefficients $([1,1]$ and $[2,2])$ shift to higher undercoolings than the $[0,0]$ curve. The $[-1,-1]$ curve also shifts similarly but to a lesser extent. This shift in $\Delta T$ is twice the order of magnitude $(10^{-2})$ than that of the previous case.  Between the two regimes, oscillatory and non-oscillatory, there is a significant difference in how convection influences the front undercooling. It is due to the formation and interaction of vortices along the growth front, which is discussed later. 

The $\lambda$ v/s $\Delta T$ curves for the three-phase $\alpha\beta\gamma$ pattern is plotted in Figure \ref{fig:eu3_LdT}. Even here, we see a similar trend. From the simulations, we observe that for values of $\lambda < 1.5\lambda_0$, the pattern does not show an oscillatory behavior, the curves closely follow each other, and the influence of convection on the attained undercooling is negligibly small. However, in the three-phase oscillatory regime between $1.5\lambda_0 - 2 \lambda_0$, we see slight shifts in the $\Delta T$ values of the curves corresponding to non-zero buoyancy coefficients. This shift is towards lower undercoolings regardless of the value of $\beta_{A,B}$ and is of an order $\sim 10 ^{-3}$ (Non-dimensionalized temperature units). Beyond $2\lambda_0$, however, the oscillations in the $\alpha\beta\gamma$ pattern becomes unstable and transforms into a two-phase pattern. This is observed in the undercooling reaching close to the two-phase curve marked $\alpha-\beta$ in Figure \ref{fig:eu3_LdT}. 

In both the cases of directional binary $\alpha\beta$ and ternary $\alpha\beta\gamma$ eutectic growth, natural convection does not significantly alter the spacing undercooling relationship in the non-oscillatory regime, and so the spacing of minimum undercooling $\lambda_0$  stays constant. The minimal influence of convection on $\lambda_0$ was also observed for isothermal Al-Cu eutectic solidification by Zhang et. al.\cite{Zhang2018_eu}. 

We investigate the emergence of the eutectic oscillations and the role of convection in the next section.

\subsubsection{Eutectic Oscillations}
\label{subsubsec:eu-oscii}

\textbf{I. $\alpha\beta$ oscillations}

\begin{figure}[ht]
	\centering
	
	\begin{subfigure}{.66\textwidth}
		\centering
		\includegraphics[width=\linewidth]{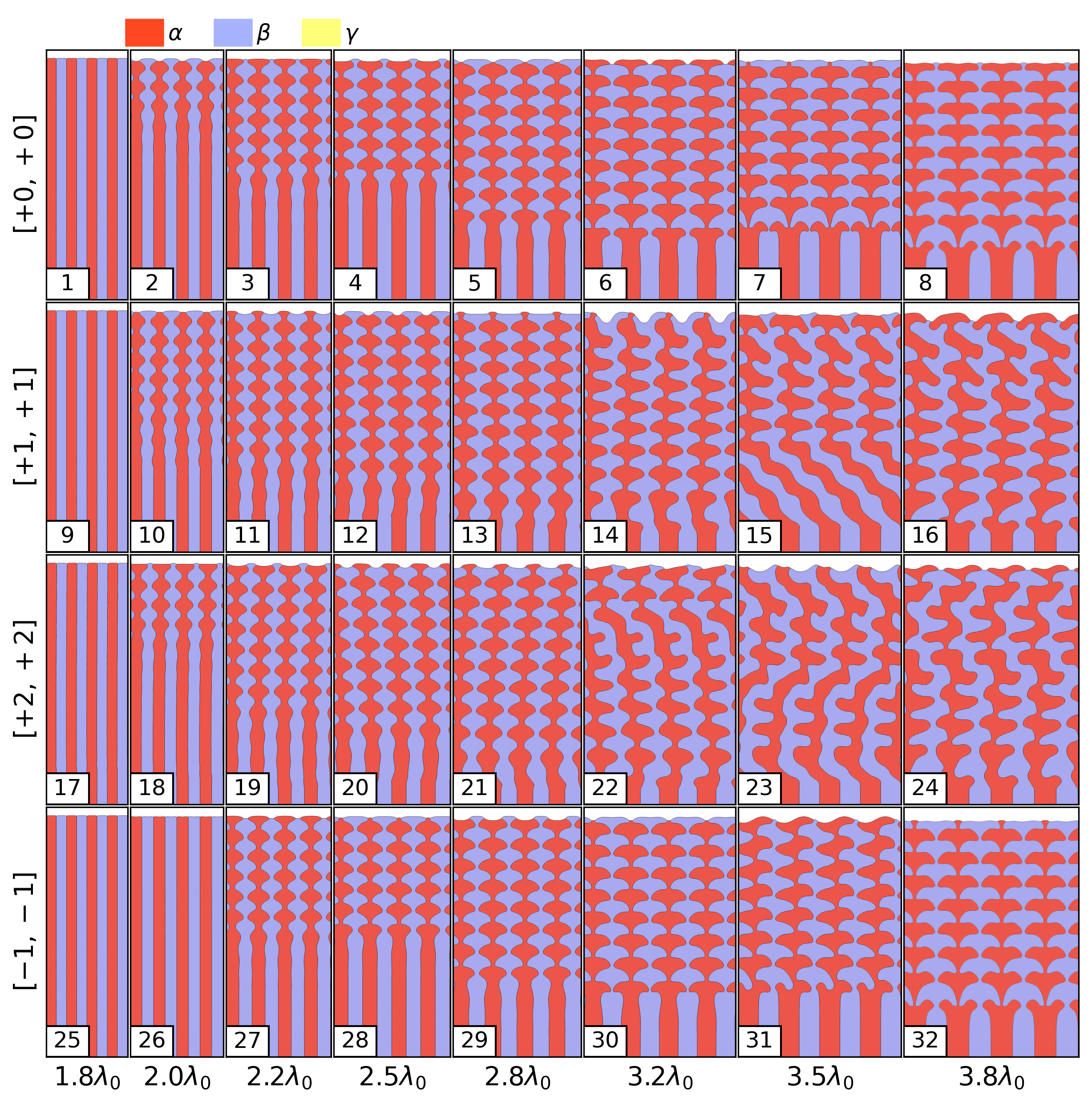}
		\caption{$\alpha\beta$ patterns.}
		\label{fig:eu_ab_unstab}
	\end{subfigure}
	\hfill
	\begin{subfigure}{.327\textwidth}
		\centering
		\includegraphics[width=\linewidth]{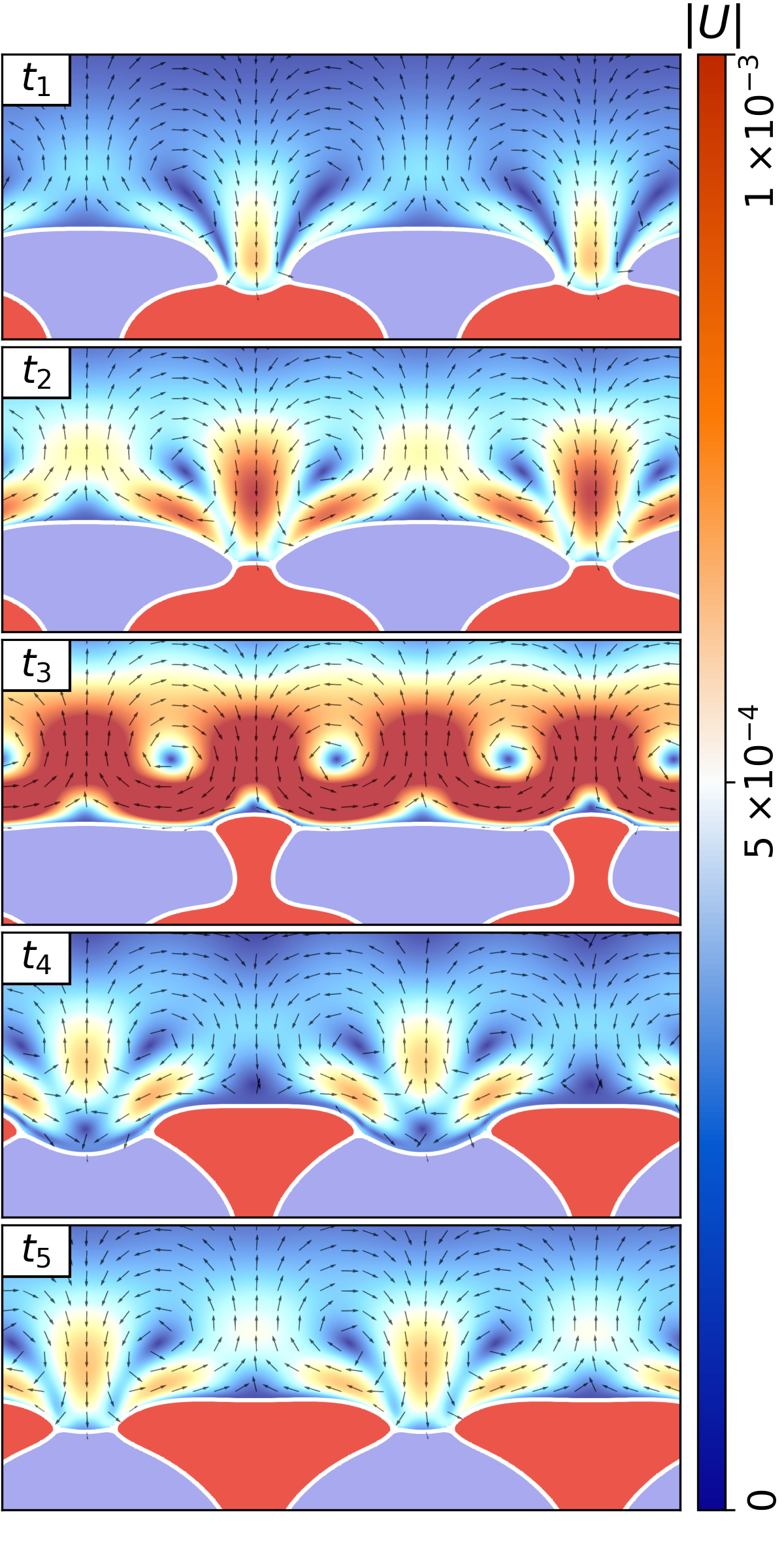}
		\caption{Velocity $\vec{U}$ fields.}
		\label{fig:eu_ab_vort}
	\end{subfigure}
	
	\caption{(a) Instabilities of the two-phase eutectic $\alpha\beta$ system under different values of $\beta_{A,B}$ (Y-axis) across multiple length scales $\lambda$ (X-axis) (b) Velocity fields associated with the two phase oscillations of pattern 28, the simulation times $t_1 < ... < t_5$, represent an oscillation half period.}
	
	\label{fig:eu3_ab}
\end{figure}

The simulated morphologies of the $\alpha\beta$ pattern at a simulation of $4\times 10^6$ are presented in Figure \ref{fig:eu_ab_unstab}. In the non oscillatory regime, we do not observe a change in steady-state width of the $\alpha$ or the $\beta$ lamella, under different buoyancy conditions. The width ratio of the phase lamella stays constant, unlike the case of isothermal eutectic growth\cite{Zhang2018_eu}.

The $\alpha\beta$ pattern shows oscillations beyond a spacing of $\lambda \ge 2\lambda_0$ (Figure \ref{fig:eu_ab_unstab}). The most common oscillation mode $(M^1_{\alpha\beta})$ is when the phases oscillate with a constant phase difference of $\pi$ and their individual oscillations are symmetric about the central vertical axis of each phase lamella. This is a commonly observed oscillation mode in binary eutectics and due to its spatial periodicity being equal to one eutectic wavelength, it is refereed to as a $1-\lambda$ oscillation\cite{Karma1996_EUTECTIC}. For the $[0,0]$ patterns $(1-8)$, this axial reflection-symmetry persists for all pattern sizes. However, for the patterns with non-zero buoyancy coefficients, it is only sustained for a lower range of pattern spacings and the threshold of this symmetry-breaking is dependent on the buoyancy coefficients. Beyond this threshold, the eutectic patterns show other symmetries and occasionally random modes of oscillations. 

Another kind of oscillation is observed prominently in the patterns 31 and 14, and briefly seen in parts of 15, 16, 22 etc. In this mode of oscillation $(M^2_{\alpha\beta})$, the central axial reflection-symmetry is lost, the phase difference is maintained, and the pattern is still an $1-\lambda$ oscillation. The defining symmetry element of the pattern in mode $M^2_{\alpha\beta}$ is a \emph{glide reflection} about the Y-axis, the glide being a translation of $\lambda_v/2$ on Y-axis ($\lambda_v$ being the vertical wavelength of the oscillation). To the authors best knowledge, this mode of oscillation has not been observed in diffusion controlled regimes of binary eutectics.

Another kind of oscillation is observed in the patterns 19-21, 14-16 etc. Here we observe that the pattern grows at an angle to the vertical axis while maintaining the the reflection-symmetric oscillation mode $M^1_{\alpha\beta}$. This tilt is attributed to the phenomenon that, in a box with periodic boundary conditions (in X-axis), convection vortices interact to form a temporary stream, leading to a non-zero mean horizontal velocity ahead of the growth front, which causes the pattern to tilt. As this is a dynamic phenomenon, the direction of the stream is chosen randomly by the system and hence the eutectic patterns are observed to tilt in either direction. Similar, tilt instabilities are also observed in diffusion controlled eutectic growth\cite{Karma1996_EUTECTIC}.

For larger spacings $\lambda \ge 3.2\lambda_0$ and for positive $\beta_{A,B}$ values, the formation of the flow streams and their dissipation back to vortices causes the observed oscillations in the $\alpha\beta$ morphology. For these cases, the stream vortex transformations become unstable, and the eutectic pattern enters a mode of random oscillations. Patterns 15-16, 22-24 show these oscillations. 

The velocity fields for the case of $\beta_{A,B} = [-1,-1]$ and $\lambda = 2.5\lambda_0$ (pattern 28), are shown in Figure \ref{fig:eu_ab_vort}. Note that the time stamps are ordered as $t_1 < ... < t_5$ and are at constant time intervals. We observe that the convection supports the growth of the $\alpha$ phase by two mechanisms, similar to the dendritic case, first flow brings in far-field liquid to the solidification front and secondly it takes away liquid ahead of the front to other regions, here phase $\beta$. This causes $\alpha$ to grow and $\beta$ to shrink. The same cycle is followed for the $\beta$ phase from $t_5$ onward.

\newpage

\begin{figure}[!ht]
	\centering
	\begin{subfigure}{.66\textwidth}
		\centering
		\includegraphics[width=\linewidth]{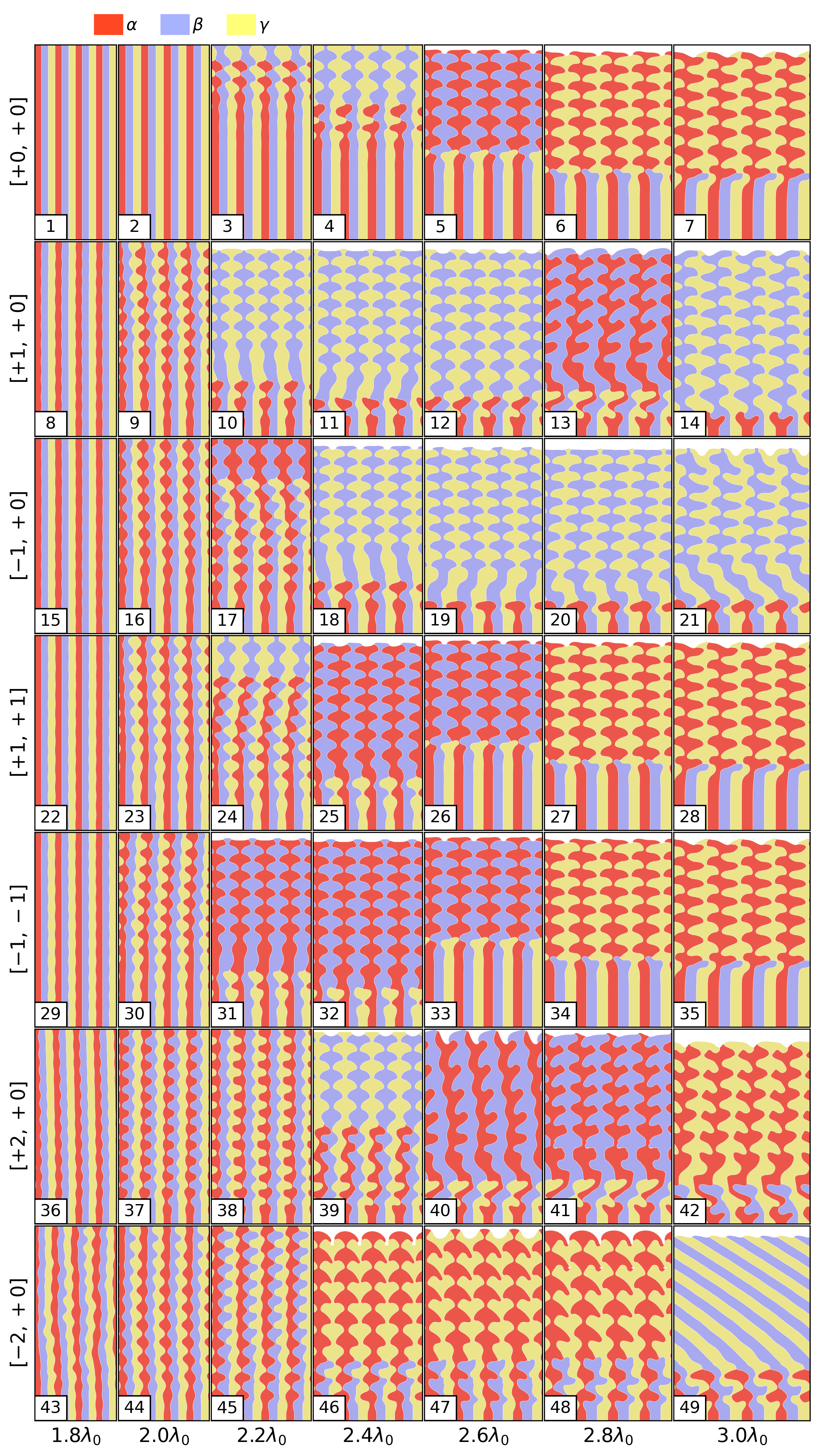}
		\caption{$\alpha\beta\gamma$ patterns.}
		\label{fig:eu_abc_unstab}
	\end{subfigure}
	\hfill
	\begin{subfigure}{.33\textwidth}
        \centering
		\includegraphics[width=\linewidth]{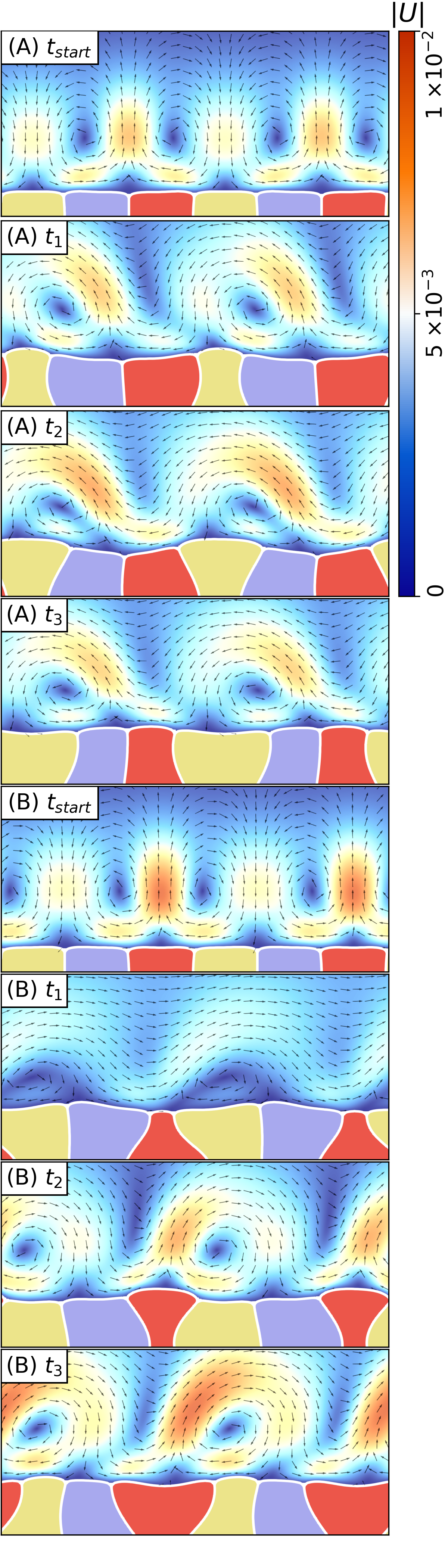}
		\caption{Velocity $\vec{U}$ fields.}
		\label{fig:eu_abc_vort}
	\end{subfigure}
	
	\caption{(a) Instabilities of the three-phase eutectic system $\alpha\beta\gamma$ under different values of $\beta_{A,B}$ (Y-axis) across multiple length scales $\lambda$ (X-axis) (b) Velocity fields associated with the three phase oscillations of (A) pattern 23 (B) pattern 9. The time stamp $t_{start}$ shows the system at early stages before the onset of oscillations. The simulation times $t_1 < ... < t_3$ represent consecutive times in an oscillation half-period. The box width is $\lambda_0$.}
	
	\label{fig:eu3_abc}
\end{figure}

{\vspace{1em}\noindent 
\textbf{II. $\alpha\beta\gamma$ oscillations}}

Similar to the $\alpha\beta$ case, in the non-oscillatory regime, the widths of the individual phase lamella in the $\alpha\beta\gamma$ eutectic system are not altered by convection. The solidification morphologies of the three-phase $\alpha\beta\gamma$ system, in the oscillatory-regime, is shown in Figure \ref{fig:eu_abc_unstab}, for different values of buoyancy coefficients (Y-axis) and pattern spacings (X-axis). 

In the Figure \ref{fig:eu_abc_unstab}, sustained three-phase oscillations are observed for non-zero $\beta_{A,B}$ and $\lambda \le 2.2\lambda_0$. The three-phase oscillation modes can be categorized as such. In type $M^1_{\alpha\beta\gamma}$, each phase lamella oscillates by changing its width at some constant phase difference w.r.t the adjacent lamella. The pattern oscillations of this type are only translationally symmetric. The top section of pattern no. 45 is a good representative of this type of three-phase oscillation. 

In oscillation mode of type $M^2_{\alpha\beta\gamma}$, one of the phases grow without a significant change in its width along the solidification direction. As the other two phases oscillate, it is seen to grow in a zig-zag like motion. This is observed in the upper section of pattern no. 37 and 38. In pattern 38 the $\beta$ (blue) phase grows at a constant width around the oscillating $\alpha, \beta$ phases, and in pattern 27 it is the $\gamma$ (yellow) phase. 

The third and the most common type of three-phase oscillations observed, type $M^3_{\alpha\beta\gamma}$, has two adjacent phase lamella oscillate collectively as one composite lamella, while maintaining a constant phase difference with the third phase lamella. A prime example is pattern no. 23, where the $\alpha,\beta$ (red,blue) phases form a composite lamella and oscillate with a phase difference of $\pi$ w.r.t $\gamma$ (yellow) phase. If we consider the composite lamella to be a distinct phase in itself, the pattern then shows the central-reflection-symmetry associated with two phase oscillations. For asymmetric values of $\beta_{A,B}$, i.e $[\pm1, 0]$ and $[\pm2,0]$, the composite lamella is always formed by the phases $\beta$ (blue) and $\gamma$ (yellow), and the $\alpha$ (red) phase oscillates independently. For the patterns at a symmetric value of $\beta_{A,B}$, i.e. $[1,1]$ and $[-1,-1]$, the composite lamella is always formed by the phases $\alpha$ and $\beta$.

The three-phase oscillation modes $M^2_{\alpha\beta\gamma}$ and $M^3_{\alpha\beta\gamma}$ are also observed in diffusion controlled growth regimes\cite{Choudhury2011_EUTECTIC}. But the exact symmetry elements of the modes are altered by convection. For instance, the central interface of the composite lamella, seen in mode $M^3_{\alpha\beta\gamma}$, is observed to oscillate under the influence of convection, unlike in a purely diffusion controlled growth regime.

Following the common trend, the development of three-phase oscillations beyond a length scale leads to the formation of two-phase structures, as seen in Figure \ref{fig:eu_abc_unstab}. Here we see a few new two-phase morphologies not observed in the earlier ranges of the buoyancy coefficients in Figure \ref{fig:eu_ab_unstab}. The two-phase oscillations of patterns 13, 40-41, does not follow the reflection-symmetry of the oscillation mode $M^1_{\alpha\beta}$ or the glide-reflection symmetry of mode $M^2_{\alpha\beta}$. In this mode $M^3_{\alpha\beta}$, the phases have completely lost their reflection symmetry and each phase oscillates with a different profile. The pattern is now only translationally symmetric with itself. This mode $M^3_{\alpha\beta}$ also characterizes the $\alpha\gamma$ oscillations of patters no. 42 and 47-48. The patterns 41 and 42 are reflections of each other, with the $\beta$ phase of 41 behaving equivalently to the $\gamma$ phase of 42. Pattern no. 46 shows us the transformation of a pattern that originally followed the $M^1_{\alpha\beta}$ symmetry but is changing to the $M^3_{\alpha\beta}$ mode as it grows. 

The velocity fields associated with two three-phase oscillation cases are shown in Figure \ref{fig:eu_abc_vort}. The time stamp $t_{start}$ shows the velocity profiles at the early stages of the simulation when the oscillations have not yet started. Case (A) is for $\beta_{A,B} = [1,1]$, pattern 23 and case (B) is for $\beta_{A,B} = [1,0]$, pattern 9. Both cases show the three phase oscillations of mode $M^3_{\alpha\beta\gamma}$, where a composite lamella exists at the initial stages of growth. We observe a correlation between the identity of the composite lamella and the flow vortices observed for the cases at time $t_{start}$. 

We see in Figure \ref{fig:eu_abc_vort} case (A), at $t_{start}$, that the $\alpha\beta$ interface lies in the symmetry-line of two alternately rotating vortices and hence it experiences zero horizontal flow ahead of it. That is not the case for the other interfaces $(\beta\gamma, \alpha\gamma)$ , which experience a net non-zero horizontal flow ahead of them. This observed flow profile limits the oscillations of the $\alpha\beta$ interface, thus forming the $\alpha\beta$ composite lamella. A similar observation can be made for the $\gamma\beta$ interface in case (B). As time progresses, the nature of flow ahead of the front changes. In later times of both cases ($t_1 ... t_3$), a dominant vortex type, clockwise in (A) and anticlockwise in (B), is seen to emerge. The nature of this vortex configuration is seen to determine the oscillation characteristics of the front. 

{\vspace{1em}\noindent 
\textbf{III. $\alpha\beta\alpha\gamma$ oscillations}}

\begin{figure}
\centering
\includegraphics[width=0.48\linewidth]{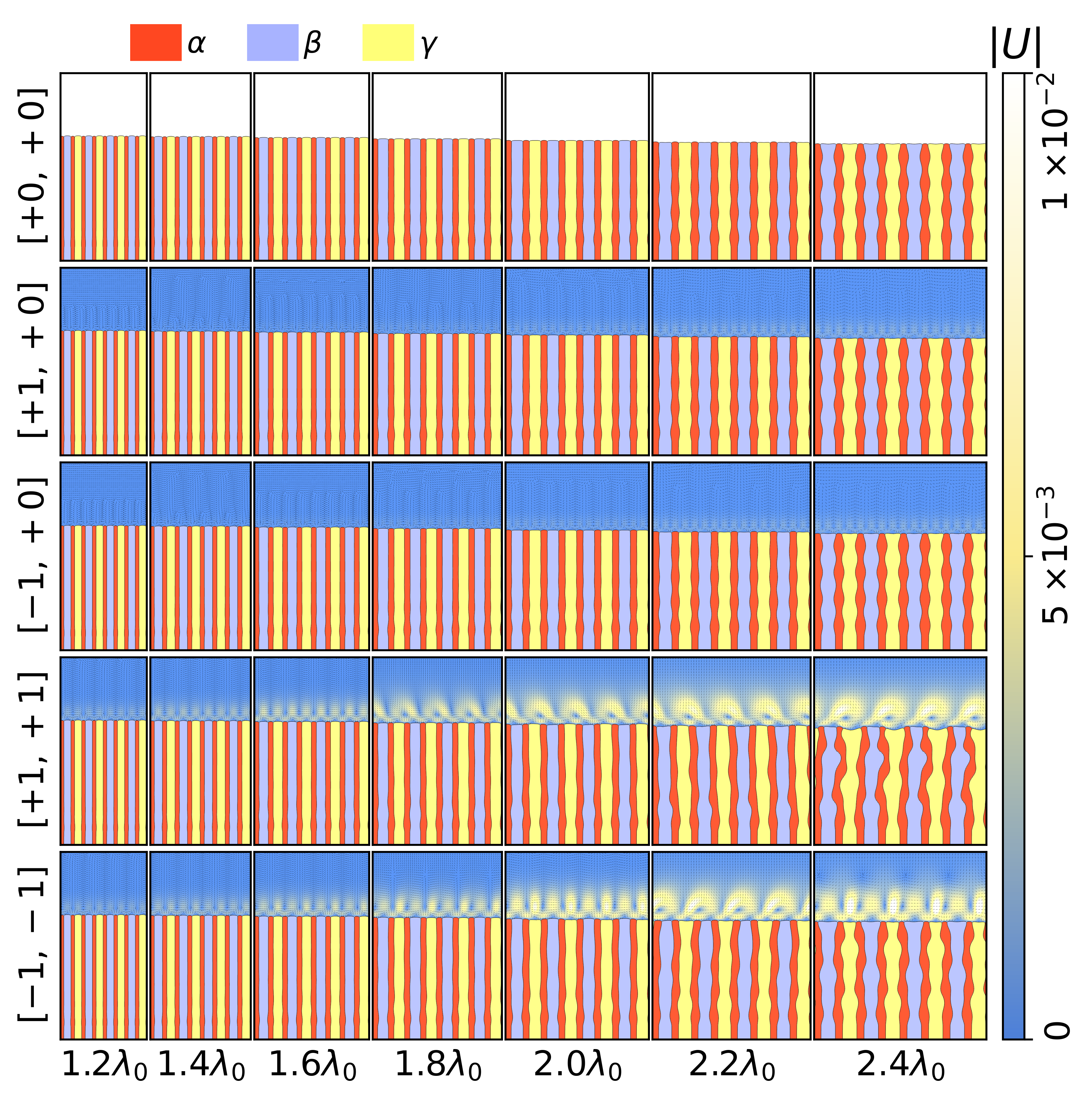}
\caption{The eutectic front and velocity maps at a simulation time of $2\times10^6 \Delta t$ for the $\alpha\beta\alpha\gamma$ pattern, at different length scales and buoyancy coefficients. The values of $[\beta_A, \beta_B]$ are on the Y-axis and the size of the pattern on X-axis.}
\label{fig:eu_abac_stab}
\end{figure}

The solidification morphologies of the $\alpha\beta\alpha\gamma$ patterns, pertaining to different points of $\lambda$ and $\beta_{A,B}$, along with the liquid velocity maps, are shown in the Figure \ref{fig:eu_abac_stab}. First we observe that the mean velocity magnitude $(|U|)$ increases with the pattern size $(\lambda)$. This phenomenon is also true for other patterns though their velocity profiles were not shown. The vortices are formed due to periodic depletion/enrichment in the concentration fields $c_A, c_B$ along the solidification front, and hence the periodicity in the vortices is inherited from the underlying eutectic pattern. Additionally, the buoyancy force exerted on the liquid ahead of the front depends on the extent of this depletion/enrichment in the concentration fields and the size of the compositional boundary layer, both of which are proportional to the diffusion length in our system, $l_D = 2D/V$, which is constant. When the pattern size is relatively small (small $\lambda/l_D$), the vortices ahead of the front are also small and are strongly affected by the friction from the drag term and fluid viscosity, so their kinetic energy dissipates. As the pattern size increases, so do the vortex sizes and the buoyancy forces across the boundary layer. This leads to the observed increase in mean velocity magnitude. The exact relationships and dependencies will be further investigated in upcoming work.

Similarly to the $\alpha\beta$ and the $\alpha\beta\gamma$ patterns, the the $\alpha\beta\alpha\gamma$ pattern shows oscillations for the upper range of spacings, shown in Figure \ref{fig:eu_abac_stab}. There are three distinct modes of oscillations observed here. The first, $M_{\alpha\beta\alpha\gamma}^1$, seen in the $[0,0]$ and $[\pm1,0]$ cases, all the phase lamella oscillate symmetrically about their central vertical axis, and the line of reflection-symmetry runs through the lamella of phases $\beta$ and $\gamma$. Second $M_{\alpha\beta\alpha\gamma}^2$, observed for $\lambda < 2.4\lambda_0$ and $\beta_{A,B} \in \{ [-1,-1] ,[+1,+1]\}$, the pattern oscillates retaining the reflection-symmetry about the $\gamma$ and $\beta$ phase lamella, but the $\alpha$ phase oscillates while maintaining its width. For $\lambda = 2.4\lambda_0$ however, the $\alpha$ phase lamella also oscillates by changing its width along the growth direction, $M_{\alpha\beta\alpha\gamma}^3$. All these three modes of oscillation are also observed in a purely diffusive eutectic growth regime \cite{Choudhury2011_EUTECTIC}. $M_{\alpha\beta\alpha\gamma}^1$ is observed at the eutectic composition whereas $M_{\alpha\beta\alpha\gamma}^2$ and $M_{\alpha\beta\alpha\gamma}^3$ are observed at off-eutectic compositions.

\section{Conclusions}
\label{sec:conclusion}

In this study, we have presented the formulation for coupling grand potential-based phase-field with a lattice-Boltzmann method to study the influence of melt-convection during dendritic and eutectic solidification. We also integrated the effects of drag, buoyancy, and shrinkage, which makes the framework robust for seamlessly modeling solidification of multi-phase multi-component systems. The proposed model is implemented using Message Passing Interface (MPI) for parallel CPU processing to expedite computation. We tested the model's viability in reproducing known results and investigated newly observed phenomenon.

During isothermal solidification, the dendrite tips favorably oriented w.r.t gravity, experience a net flow towards them. This increases the flux at the tips and cause the tip radius and velocity to increase. For dendrite arms unfavorably oriented w.r.t. gravity, convection causes tip splitting and forms solute plumes. 

During directional solidification, no morphological instability was found to exist for the case of negative buoyancy coefficients, i.e. when the inter-dendritic liquid is heavier than the for field liquid. Additionally, the solidification front grows stably even for small positive values of the buoyancy coefficient. In this stable regime, buoyancy conditions that lead to flow towards the dendrite tips, caused the tip radii and PDAS to increase by increasing the solute flux and vice versa. During the stable growth regime two alternately rotating vortices were associated with two-adjacent dendrite tips. As the buoyancy forces increase, the two vortex configuration changes to a configuration of one vortex between two adjacent tips. This causes preferential enrichment and depletion of fluxes across alternate dendrites and causes the stable solidification front to oscillate. As the buoyancy forces increase further these oscillations increase in their amplitude and become unstable. This instability can be suppressed by either increasing the pulling velocity or tilting the seed orientation.

During eutectic solidification, at pattern sizes significantly larger than the size of minimum undercooling, there exists eutectic oscillations. We have identified certain new modes of oscillations which were not observed in a purely diffusive growth regime. The eutectic oscillations are due to the change of the local composition gradients ahead of the solidification front caused by natural-convection. We have qualitatively analyzed the oscillation modes, but investigating their direct relationship to the buoyancy coefficients remains as a future work.

In conclusion, we have demonstrated the effectiveness of the Grand-Potential Phase-Field Lattice-Boltzmann formulation in simulating the solidification process along with convection and have shown its capability to investigate and discover convection induced morphological instabilities in multi-phase and multi-component systems.

\section*{Acknowledgments}
The partial financial support received from the Defense Research Organization (DRDO), Government of India is gratefully acknowledged. 
\clearpage
\appendix
\section{Computational scheme}
\label{app:comScheme}

This section presents a brief details on the computational scheme and material parameters used in the simulation. We solve coupled Phase-Field Lattice-Boltzmann model detailed in Section \ref{sec:phase-field_LBM} in staggered manner. We use the finite difference scheme to discretize the grand potential-based phase-field and diffusion equation given in Eqs. \ref{eq:PhfGrand} and \ref{eq:conc}, respectively. The lattice Boltzmann equation for multicomponent alloys given in Eq.\ref{eq:LBMEq_1} is discretized in velocity space in two and three dimensions using the D2Q9 and D3Q27 lattice velocity models respectively. Their weight coefficients and lattice velocities are mentioned in Table \ref{table:LBM_Schemes} . Note that the LBM weighting factors must satisfy unity sum, symmetry (independent of the rotation of the lattices), and isotropy conditions\cite{mohamad2011lattice}. For spatial discretization of the LBM equations, we use same mesh used for the phase-field discretization. 

\begin{figure}
\centering
\begin{tabular}{cc}

\begin{tikzpicture}[scale=0.8]
    \draw[gray, dashed] (0,0) rectangle (4,4);

    \draw[thick,-{>[scale=1.5]}] (2,2) -- (4,2) node[midway, below] {}; 
    \draw[thick,-{>[scale=1.5]}] (2,2) -- (2,4) node[midway, left] {}; 
    \draw[thick,-{>[scale=1.5]}] (2,2) -- (0,2) node[midway, below] {}; 
    \draw[thick,-{>[scale=1.5]}] (2,2) -- (2,0) node[midway, right] {}; 

    \draw[thick,-{>[scale=1.5]}] (2,2) -- (4,4) node[midway, above left] {};
    \draw[thick,-{>[scale=1.5]}] (2,2) -- (0,4) node[midway, above right] {};
    \draw[thick,-{>[scale=1.5]}] (2,2) -- (0,0) node[midway, below right] {};
    \draw[thick,-{>[scale=1.5]}] (2,2) -- (4,0) node[midway, below left] {};

    \node[left] at (0,0) {\textbf{7}};
    \node[below] at (2,0) {\textbf{4}};
    \node[right] at (4,0) {\textbf{8}};
    \node[right] at (4,2) {\textbf{1}};
    \node[right] at (4,4) {\textbf{5}};
    \node[above] at (2,4) {\textbf{2}};
    \node[left] at (0,4) {\textbf{6}};
    \node[left] at (0,2) {\textbf{3}};
    \node at (2.4, 2.2) {\textbf{0}};
\end{tikzpicture}  &

\begin{tikzpicture}[scale=0.9]
    \draw[thick, dashed] (0,0,0) -- (4,0,0);
    \draw[thick, dashed] (0,4,0) -- (4,4,0);
    \draw[thick, dashed] (0,0,4) -- (4,0,4);
    \draw[thick, dashed] (0,4,4) -- (4,4,4);
    \draw[thick, dashed] (0,0,0) -- (0,4,0);
    \draw[thick, dashed] (4,0,0) -- (4,4,0);
    \draw[thick, dashed] (0,0,4) -- (0,4,4);
    \draw[thick, dashed] (4,0,4) -- (4,4,4);
    \draw[thick, dashed] (0,0,0) -- (0,0,4);
    \draw[thick, dashed] (4,0,0) -- (4,0,4);
    \draw[thick, dashed] (0,4,0) -- (0,4,4);
    \draw[thick, dashed] (4,4,0) -- (4,4,4);
    \draw[thick, dotted] (2,0,0) -- (2,4,0);
    \draw[thick, dotted] (0,2,0) -- (4,2,0);
    \draw[thick, dotted] (2,0,4) -- (2,4,4);
    \draw[thick, dotted] (0,2,4) -- (4,2,4);
    \draw[thick, dotted] (0,0,2) -- (4,0,2);
    \draw[thick, dotted] (2,0,0) -- (2,0,4);
    \draw[thick, dotted] (0,4,2) -- (4,4,2);
    \draw[thick, dotted] (2,4,0) -- (2,4,4);
    \draw[thick, dotted] (0,0,2) -- (0,4,2);
    \draw[thick, dotted] (0,2,0) -- (0,2,4);
    \draw[thick, dotted] (4,0,2) -- (4,4,2);
    \draw[thick, dotted] (4,2,0) -- (4,2,4);

    \coordinate (center) at (2,2,2);
    \foreach \x/\y/\z/\label in {
        0/0/0/20, 4/0/0/26, 0/4/0/24, 4/4/0/21, 
        0/0/4/22, 4/0/4/23, 0/4/4/25, 4/4/4/19,
        2/0/0/12, 2/4/0/17, 2/0/4/18, 2/4/4/11, 
        0/2/0/10, 4/2/0/15, 0/2/4/16, 4/2/4/9,
        0/0/2/8, 4/0/2/13, 0/4/2/14, 4/4/2/7,
        2/2/0/6, 2/2/4/5, 2/0/2/4, 2/4/2/3, 
        0/2/2/1, 4/2/2/2
    } {
        \coordinate (v\x\y\z) at (\x,\y,\z);
        \draw[-{{>[scale=1.5]}[scale=1.5]}, thick] (center) -- (v\x\y\z);
    }
    \foreach \x/\y/\z/\label in {
        0/0/0/20,
        0/4/0/24,
        0/0/4/22,
        0/4/4/25,
        0/2/0/10,
        0/2/4/16,
        0/0/2/8,
        0/4/2/14,
        0/2/2/2
    } {
        \coordinate (v\x\y\z) at (\x,\y,\z);
        \node[left] at (v\x\y\z) {\textbf{\label}};
    }
    \foreach \x/\y/\z/\label in {
        4/0/0/26,
        4/4/0/21,
        4/0/4/23,
        4/4/4/19,
        4/2/0/15,
        4/2/4/9,
        4/0/2/13,
        4/4/2/7,
        4/2/2/1
    } {
        \coordinate (v\x\y\z) at (\x,\y,\z);
        \node[right] at (v\x\y\z) {\textbf{\label}};
    }
    \node[above] at (2,4,0) {\textbf{17}};
    \node[above] at (2,4,4) {\textbf{11}};
    \node[above] at (2,4,2) {\textbf{3}};
    \node[below] at (2,0,0) {\textbf{12}};
    \node[below] at (2,2,0) {\textbf{6}};
    \node[below] at (2,2,4) {\textbf{5}};
    \node[below] at (2,0,4) {\textbf{18}};
    \node[below] at (2,0,2) {\textbf{4}};
    \node[below] at (1.6,2,2) {\textbf{0}};
    \end{tikzpicture}
\end{tabular}
\caption{D2Q9(left) and D3Q27(right) lattice used in lattice Boltzmann method.}
\label{fig:lbmcvec}
\end{figure}
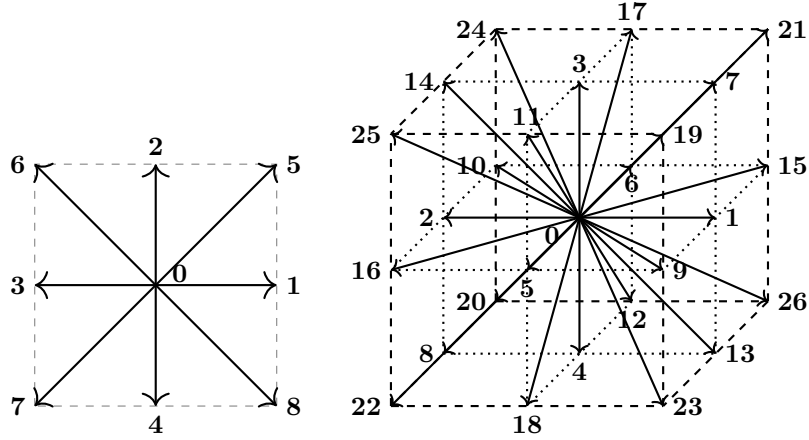




\begin{table}
    \caption{Lattice velocity sets and weight coefficients for D2Q9 and D3Q27 discretization schemes.}
    \label{table:LBM_Schemes}
    \centering
    \begin{tabular}{|llccc|}
        \hline 
        Scheme & Index $k$ & Velocities $c_k$ & Length $|c_k|$ & Weight $w_k$   \\
        \hline
        D2Q9 & $k=0$ & $(0,0)$ & $0$ & 4/9 \\
        & $k=1-4$ & $ (\pm 1, 0), (0, \pm 1) $ & $1$  & 1/9 \\
        & $k=5-8$ & $ (\pm 1, \pm 1) $ & $\sqrt{2}$  & 1/36 \\
        

        D3Q27 & $k=0$ & $(0,0,0)$ & $0$ & 8/27 \\
        & $k=1-6$ & $(\pm 1,0,0), (0,\pm 1,0), (0,0,\pm 1)$&  $1$ & 2/27 \\
        & $k=7-18$ & $(\pm 1, \pm 1,0), (0,\pm 1, \pm 1), (\pm 1,0,\pm 1)$ & $\sqrt{2}$ & 1/54 \\
        & $k=19-26$ & $(\pm 1, \pm 1,\pm 1)$ & $\sqrt{3}$ & 1/216 \\
        \hline
    \end{tabular}
\end{table}

Chapman–Enskog analysis demonstrates that the lattice Boltzmann (LB) equation is equivalent to the Navier-Stokes (NS) equation with an accuracy up to second-order precision in the Knudsen number $(Ku)$, i.e., $\mathcal{O}(Ku^2)$, when the lattice relaxation time $(\tau_{LBM})$ is appropriately selected. It can be expressed in terms of the viscosity $\nu_{LBM}$ parameter as follows:

\begin{equation}
\nu_{LBM} = c_s^2 \left(\tau_{LBM} - \frac{1}{2} \right) \frac{(\delta x)^2}{\delta t},
\end{equation}

where $c_s = 1/\sqrt{3}$ for both the lattices D2Q9 and D3Q27. We use the same grid points for the computation of finite differences in the phase field and LBM. The grids are uniform, and we also use the same temporal discretization. For ease of numerical implementation, we may perform nondimensional analysis and convert the physical quantities into LB units and vice versa. We use the concept of similarity and scaling to transfer the units from the physical space to the lattice space. It is common practice and recommended to set the lattice size $\delta x$ and $\delta t$ as unity, and $\rho^* = 1$. The superscript $^*$ denotes the quantities in LB units. So, the conversion factor for length $(C_l)$, time $(C_t)$, and density/mass $(C_\rho)$ is:

\begin{equation}
C_l = \frac{1}{\Delta x}, 
C_t = \frac{1}{\Delta t},
C_\rho = \frac{\rho}{\rho^*}  C_l^3.
\end{equation}

where $\Delta x$ and $\Delta t$ are in physical units. Similarly, we may derive the conversion factors for velocity and viscosity as:

\begin{equation}
\frac{\tb{U}}{\tb{\: U}^*} = \frac{C_l}{C_t}, 
\frac{\nu}{\nu^*} = \frac{C_l^2}{C_t}.
\end{equation}

The staggered solution scheme is presented in Algorithm \ref{algo:phfLBM}. To expedite computation, we developed the coupled LBM-phase-field code using C and the Message Passing Interface (MPI) for parallel CPU processing. The MPI code for the grand potential-based phase-field model is available at \href{github.com/ICME-India/MicroSim/tree/main/Grand_potential_Finite_difference_2D_MPI}{MicroSim-GitHub}. as open source.

\begin{algorithm}
\caption{Coupled Phase-Field and Lattice-Boltzmann Method for Dendritic Solidification}
\begin{algorithmic}[1]
    \Statex \textbf{Input:} Initial liquid concentration, phase-field $( \phi)$, material properties, lattice parameters
    \Statex \textbf{Output:} Simulation results including phase distribution and fluid flow  \vspace{3mm}  
    \Statex \textbf{Initialization:} Initialize phase-field variables, liquid concentration, fluid velocity, LB distribution function, and set the  boundary conditions 
    
    \While{$n_{step} \leq N_{total}$}    \Comment{loop over all the time steps}
        \State Solve for phase-field variable $(\phi)$         \Comment{Eq. \ref{eq:PhfGrand}}
        \Statex \hspace{2em} take $\Tilde{\mu}_i$ from the previous step

        \State Solve the chemical potential equation        \Comment{Eq. \ref{eq:conc}}
        \Statex \hspace{2em} use $\tb{U}$ from the previous step and phase-field from the current step

          
        \State Solve the lattice-Boltzmann equations \Comment{see, $\S$ \ref{subsec:LBM}}
            \Statex \hspace{2em}  Compute the equilibrium distribution function    \Comment{use Eq. \ref{eq:f^eq}}
            \Statex \hspace{2em}  Compute the discrete forcing $G_k^i$ and update $f^i_k$    \Comment{use Eq. \ref{eq:discreteForce}}
            \Statex \hspace{2em}  Collision:
            \Statex \hspace{2em} \hspace{2em} \(  f_k^{i*}(\bs{x}, t)  \leftarrow f^i_k(\bs{x}, t) \left(1-\frac{\Delta t}{\tau} \right) + f_{i, k}^{eq}(\bs{x}, t) \frac{\Delta t}{\tau} \) 
            \Statex \hspace{2em}  Streaming: 
            \Statex \hspace{2em} \hspace{2em} \( f^i_k(\bs{x} + \bs{c}_k  \Delta t, t +\Delta t )  \leftarrow f_k^{i*}(\bs{x}, t)\)
            \Statex \hspace{2em}  Apply boundary conditions and update the LB distribution functions
            \Statex \hspace{2em}  Momentum:   \Comment{use Eq. \ref{eq:rhovel}}
            \Statex \hspace{2em} \hspace{2em} \( f^i_k(\bs{x} + \bs{c}_k  \Delta t, t +\Delta t ) \rightarrow  \rho_i, \tb{U}\)
           
        \State \textbf{Write output:}
        \Statex \hspace{2em}  $\rho, \phi, \tb{U}, C  \rightarrow \text{disk} $ 
        
    \EndWhile
\end{algorithmic}
\label{algo:phfLBM}
\end{algorithm}


\section{Simulation Parameters}
\label{app:simpara}
The universal gas constant, $R$, and the molar volume, $V$, are both normalized to unity. Similarly, all solid--solid and solid--liquid interface energies, $\bm{\gamma}_{\alpha\beta}$, are assigned a value of $1.0$, thereby eliminating the influence of anisotropic interfacial energies.

Diffusion is assumed to occur exclusively in the liquid phase. Accordingly, the diffusivity matrices of all solid phases are taken to be identically zero,
$$
\bm{D}_{ij}^{\alpha}=\mathbf{0},
$$
while the liquid diffusivity matrix, $\bm{D}_{ij}^{l}$, is assumed to be diagonal (cross-diffusion effects are neglected). Each diagonal entry of $\bm{D}_{ij}^{l}$ is assigned a value of $1.0$, resulting in equal diffusivities for all independent chemical components in the liquid.

The equilibrium temperature, $T_{eq}$, is normalized to $1.0$. Unless explicitly stated, the lattice-Boltzmann and phase-field parameters are held fixed throughout all simulations. The lattice-Boltzmann kinematic viscosity is set to $\nu_{LBM}=0.025$, corresponding to a relaxation time of $\tau_{LBM}=0.027$. The physical kinematic viscosity is prescribed as $\nu=0.97$, while a uniform body force of magnitude $\bm{g}=0.01$ is applied to drive fluid motion. The diffuse interface thickness is chosen as $W_0=0.8$. 

\subsection{Binary Solidification}
\begin{table}
    \caption{Simulation parameters (non-dimensional units) for the binary solidification simulations.}
    \label{table:simpara}
    \centering
    \begin{tabular}{|lll|}
        \hline
        Parameter & Description & Value \\
        \hline
        $dx$        & Mesh resolution                  & 12.0 \\
        $\epsilon$  & Diffuse interface width          & 48.0 \\
        $dt$        & Time step                        & 2.88 \\
        $f_{s-l}$       & Solid--liquid anisotropy fold      & 4 \\
        $\delta_{s-l}$  & Solid--liquid anisotropy strength  & 0.02 \\
        $T_{sim}$   & Isothermal simulation temperature & 0.92 \\
        $V_p$       & Pulling velocity                 & $1.0\times10^{-2}$ \\
        $G$         & Thermal gradient                 & $0.12/19200$ \\
        \hline
    \end{tabular}
\end{table}

The numerical parameters specific to the binary dendritic solidification simulations are summarized in Table~\ref{table:simpara}. Unless stated otherwise, these values are kept fixed throughout the study. 

A four-fold anisotropy ($f_{s-l}=4$) with anisotropy strength $\delta_{s-l}=0.02$ is prescribed for the solid-liquid interfacial energy, thereby promoting the formation of dendritic morphologies characteristic of cubic crystalline materials. For isothermal solidification (Case 1) simulations, the entire computational domain is maintained at a constant temperature of $T_{sim}=0.92$. In contrast, directional solidification (Case 2) simulations are performed by imposing a constant pulling velocity, $V_p$, together with a prescribed thermal gradient, $G$. 

Unless explicitly stated, the densities of the two phases are assumed to be constant, with the solid and liquid densities are taken as $\rho_{\mathrm{solid}} = 1.2, \rho_{\mathrm{liq}} = 1.0.$

\subsection{Eutectic Solidification}

The numerical parameters used for the eutectic solidification simulations are summarized in Table~\ref{table:simparaeu3}.  The interface energies between all phases are assumed to be isotropic. The densities of the solid and liquid phases are prescribed as $\rho_{\mathrm{solid}} = 1.2, \rho_{\mathrm{liq}} = 1.0.$

\begin{table}[!ht]
    \caption{Simulation parameters (non-dimensional units) for the eutectic solidification simulations.}
    \label{table:simparaeu3}
    \centering
    \begin{tabular}{|lll|}
        \hline
        Parameter & Description & Value \\
        \hline
        $dx$        & Mesh resolution         & 5.0 \\
        $\epsilon$  & Diffuse interface width & 20.0 \\
        $dt$        & Time step               & 1.0 \\
        $V_p$       & Pulling velocity        & $2.0\times10^{-3}$ \\
        $G$         & Thermal gradient        & $0.04/800$ \\
        \hline
    \end{tabular}
\end{table}

The thermodynamic equilibrium data required by the phase-field model are specified through the equilibrium compositions of each phase and the corresponding liquid phase. The equilibrium compositions are chosen as
$$
c^{eq}_{\alpha-\alpha} =
\left[0.6,\;0.2\right], \qquad
c^{eq}_{\beta-\beta} =
\left[0.2,\;0.6\right], \qquad
c^{eq}_{\gamma-\gamma} =
\left[0.2,\;0.2\right],\qquad
c^{eq}_{l-l} =
\left[1/3,\;1/3\right].
$$

The slopes of the coexistence lines are prescribed independently for the solid-state and solid--liquid equilibria. The slopes corresponding to the solid-state equilibrium are given by
$$
m^{eq}_{\alpha-\alpha} =
\left[+1.6,\;0.0\right], \qquad
m^{eq}_{\beta-\beta} =
\left[0.0,\;+1.6\right], \qquad
m^{eq}_{\gamma-\gamma} =
\left[-1.6,\;-1.6\right],
$$
whereas the solid--liquid coexistence lines are defined by
$$
m^{eq}_{\alpha-l} =
\left[+1.2,\;0.0\right], \qquad
m^{eq}_{\beta-l} =
\left[0.0,\;+1.2\right], \qquad
m^{eq}_{\gamma-l} =
\left[-1.2,\;-1.2\right].
$$

\clearpage
\bibliography{ref_gen, ref_exp, ref_sim}

@article{Spinelli2004,
author = {Spinelli, J. E. and Ferreira, I. L. and Garcia, A.},
doi = {10.1016/j.jallcom.2004.04.098},
issn = {09258388},
journal = {Journal of Alloys and Compounds},
number = {1-2},
pages = {217--226},
title = {{Influence of melt convection on the columnar to equiaxed transition and microstructure of downward unsteady-state directionally solidified Sn-Pb alloys}},
volume = {384},
year = {2004}
}

@article{Liu2006,
author = {Liu, S. and Li, J. and Lee, J. and Trivedi, R.},
doi = {10.1080/09500830500504036},
issn = {14786435},
journal = {Philosophical Magazine},
number = {24},
pages = {3717--3738},
title = {{Spatio-temporal microstructure evolution in directional solidification processes}},
volume = {86},
year = {2006}
}

@article{Ruvalcaba2007,
author = {Ruvalcaba, D. and Mathiesen, R. H. and Eskin, D. G. and Arnberg, L. and Katgerman, L.},
doi = {10.1016/j.actamat.2007.03.030},
issn = {13596454},
journal = {Acta Materialia},
number = {13},
pages = {4287--4292},
title = {{In situ observations of dendritic fragmentation due to local solute-enrichment during directional solidification of an aluminum alloy}},
volume = {55},
year = {2007}
}

@article{Shevchenko2013,
author = {Shevchenko, Natalia and Boden, Stephan and Gerbeth, Gunter and Eckert, Sven},
doi = {10.1007/s11661-013-1711-1},
issn = {10735623},
journal = {Metallurgical and Materials Transactions A: Physical Metallurgy and Materials Science},
number = {8},
pages = {3797--3808},
title = {{Chimney formation in solidifying Ga-25wt pct in alloys under the influence of thermosolutal melt convection}},
volume = {44},
year = {2013}
}

@article{Reinhart2013,
author = {Reinhart, G. and Gandin, Ch A. and Mangelinck-No{\"{e}}l, N. and Nguyen-Thi, H. and Spinelli, J. E. and Baruchel, J. and Billia, B.},
doi = {10.1016/j.actamat.2013.04.067},
issn = {13596454},
journal = {Acta Materialia},
number = {13},
pages = {4765--4777},
title = {{Influence of natural convection during upward directional solidification: A comparison between in situ X-ray radiography and direct simulation of the grain structure}},
volume = {61},
year = {2013}
}

@article{Shevchenko2015,
author = {Shevchenko, N. and Roshchupkina, O. and Sokolova, O. and Eckert, S.},
doi = {10.1016/j.jcrysgro.2014.11.043},
issn = {00220248},
journal = {Journal of Crystal Growth},
pages = {1--8},
publisher = {Elsevier},
title = {{The effect of natural and forced melt convection on dendritic solidification in Ga-In alloys}},
url = {http://dx.doi.org/10.1016/j.jcrysgro.2014.11.043},
volume = {417},
year = {2015}
}

@article{Reinhart2020,
author = {Reinhart, G. and Grange, D. and Abou-Khalil, L. and Mangelinck-No{\"{e}}l, N. and Niane, N. T. and Maguin, V. and Guillemot, G. and Gandin, Ch A. and Nguyen-Thi, H.},
doi = {10.1016/j.actamat.2020.04.003},
issn = {13596454},
journal = {Acta Materialia},
pages = {68--79},
title = {{Impact of solute flow during directional solidification of a Ni-based alloy: In-situ and real-time X-radiography}},
volume = {194},
year = {2020}
}

@book{dantzigBook,
  title={Solidification: -Revised \& Expanded},
  author={Dantzig, Jonathan A and Rappaz, Michel},
  year={2016},
  publisher={EPFL press}
}

@article{choudhury2012grand,
  title={Grand-potential formulation for multicomponent phase transformations combined with thin-interface asymptotics of the double-obstacle potential},
  author={Choudhury, Abhik and Nestler, Britta},
  journal={Physical Review E},
  volume={85},
  number={2},
  pages={021602},
  year={2012},
  publisher={APS}
}

@article{Choudhury2015,
author = {Choudhury, Abhik and Kellner, Michael and Nestler, Britta},
doi = {10.1016/j.cossms.2015.03.003},
issn = {13590286},
journal = {Current Opinion in Solid State and Materials Science},
number = {5},
pages = {287--300},
publisher = {Elsevier Ltd},
title = {{A method for coupling the phase-field model based on a grand-potential formalism to thermodynamic databases}},
url = {http://dx.doi.org/10.1016/j.cossms.2015.03.003},
volume = {19},
year = {2015}
}

@article{LBM_Chen1998,
author = {Chen, Shiyi and Doolen, Gary D.},
doi = {10.1146/annurev.fluid.30.1.329},
issn = {00664189},
journal = {Annual Review of Fluid Mechanics},
number = {Kadanoff 1986},
pages = {329--364},
title = {{Lattice boltzmann method for fluid flows}},
volume = {30},
year = {1998}
}

@book{mohamad2011lattice,
  title={Lattice boltzmann method},
  author={Mohamad, AA},
  volume={70},
  year={2011},
  publisher={Springer}
}

@article{Samanta2022,
author = {Samanta, Runa and Chattopadhyay, Himadri and Guha, Chandan},
doi = {10.1016/j.commatsci.2022.111288},
issn = {09270256},
journal = {Computational Materials Science},
number = {March},
pages = {111288},
publisher = {Elsevier B.V.},
title = {{A review on the application of lattice Boltzmann method for melting and solidification problems}},
url = {https://doi.org/10.1016/j.commatsci.2022.111288},
volume = {206},
year = {2022}
}

@article{Viardin2020,
author = {Viardin, A. and Zollinger, J. and Sturz, L. and Apel, M. and Eiken, J. and Berger, R. and Hecht, U.},
doi = {10.1016/j.commatsci.2019.109358},
issn = {09270256},
journal = {Computational Materials Science},
number = {September 2019},
title = {{Columnar dendritic solidification of TiAl under diffusive and hypergravity conditions investigated by phase-field simulations}},
volume = {172},
year = {2020}
}

@article{Selzer2009,
author = {Selzer, Michael and Jainta, Marcus and Nestler, Britta},
doi = {10.1002/pssb.200844282},
issn = {03701972},
journal = {Physica Status Solidi (B) Basic Research},
number = {6},
pages = {1197--1205},
title = {{A Lattice-Boltzmann model to simulate the growth of dendritic and eutectic microstructures under the influence of fluid flow}},
volume = {246},
year = {2009}
}

@article{Rojas2015,
author = {Rojas, Roberto and Takaki, Tomohiro and Ohno, Munekazu},
doi = {10.1016/j.jcp.2015.05.045},
issn = {10902716},
journal = {Journal of Computational Physics},
pages = {29--40},
publisher = {Elsevier Inc.},
title = {{A phase-field-lattice Boltzmann method for modeling motion and growth of a dendrite for binary alloy solidification in the presence of melt convection}},
url = {http://dx.doi.org/10.1016/j.jcp.2015.05.045},
volume = {298},
year = {2015}
}

@article{Sakane2017,
author = {Sakane, Shinji and Takaki, Tomohiro and Rojas, Roberto and Ohno, Munekazu and Shibuta, Yasushi and Shimokawabe, Takashi and Aoki, Takayuki},
doi = {10.1016/j.jcrysgro.2016.11.103},
issn = {00220248},
journal = {Journal of Crystal Growth},
number = {November 2016},
pages = {154--159},
publisher = {Elsevier B.V.},
title = {{Multi-GPUs parallel computation of dendrite growth in forced convection using the phase-field-lattice Boltzmann model}},
url = {http://dx.doi.org/10.1016/j.jcrysgro.2016.11.103},
volume = {474},
year = {2017}
}

@article{Takaki2017,
author = {Takaki, Tomohiro and Rojas, Roberto and Sakane, Shinji and Ohno, Munekazu and Shibuta, Yasushi and Shimokawabe, Takashi and Aoki, Takayuki},
doi = {10.1016/j.jcrysgro.2016.11.099},
issn = {00220248},
journal = {Journal of Crystal Growth},
number = {November 2016},
pages = {146--153},
publisher = {Elsevier B.V.},
title = {{Phase-field-lattice Boltzmann studies for dendritic growth with natural convection}},
url = {http://dx.doi.org/10.1016/j.jcrysgro.2016.11.099},
volume = {474},
year = {2017}
}

@article{Takaki2018,
author = {Takaki, Tomohiro and Sato, Ryotaro and Rojas, Roberto and Ohno, Munekazu and Shibuta, Yasushi},
doi = {10.1016/j.commatsci.2018.02.004},
issn = {09270256},
journal = {Computational Materials Science},
pages = {124--131},
publisher = {Elsevier B.V.},
title = {{Phase-field lattice Boltzmann simulations of multiple dendrite growth with motion, collision, and coalescence and subsequent grain growth}},
url = {https://doi.org/10.1016/j.commatsci.2018.02.004},
volume = {147},
year = {2018}
}

@article{Ratkai2019,
author = {R{\'{a}}tkai, L{\'{a}}szl{\'{o}} and Pusztai, Tam{\'{a}}s and Gr{\'{a}}n{\'{a}}sy, L{\'{a}}szl{\'{o}}},
doi = {10.1038/s41524-019-0250-8},
issn = {20573960},
journal = {npj Computational Materials},
number = {1},
publisher = {Springer US},
title = {{Phase-field lattice Boltzmann model for dendrites growing and moving in melt flow}},
url = {http://dx.doi.org/10.1038/s41524-019-0250-8},
volume = {5},
year = {2019}
}

@article{Takaki2020,
author = {Takaki, Tomohiro and Sakane, Shinji and Ohno, Munekazu and Shibuta, Yasushi and Aoki, Takayuki},
doi = {10.1016/j.commatsci.2019.109209},
issn = {09270256},
journal = {Computational Materials Science},
number = {September 2019},
pages = {109209},
publisher = {Elsevier},
title = {{Large–scale phase–field lattice Boltzmann study on the effects of natural convection on dendrite morphology formed during directional solidification of a binary alloy}},
url = {https://doi.org/10.1016/j.commatsci.2019.109209},
volume = {171},
year = {2020}
}

@article{Yang2019,
doi = {10.1016/j.commatsci.2018.11.024},
issn = {09270256},
journal = {Computational Materials Science},
number = {November 2018},
pages = {130--139},
publisher = {Elsevier},
title = {{Study of dendrite growth with natural convection in superalloy directional solidification via a multiphase-field-lattice Boltzmann model}},
url = {https://doi.org/10.1016/j.commatsci.2018.11.024},
volume = {158},
year = {2019}
}

@article{Zhang2022,
author = {Zhang, Shijie and Li, Chenyu and Li, Ri},
doi = {10.1016/j.mtcomm.2022.103342},
issn = {23524928},
journal = {Materials Today Communications},
number = {March},
title = {{Cellular automata-lattice Boltzmann simulation of multi-dendrite motion under convection based on dynamic grid technology}},
volume = {31},
year = {2022}
}

@article{Zhang2024,
author = {Zhang, Shijie and Zhu, Baofeng and Li, Yunbo and Zhang, Yang and Li, Ri},
doi = {10.1016/j.commatsci.2024.113308},
issn = {09270256},
journal = {Computational Materials Science},
number = {August},
title = {{Cellular automata-lattice Boltzmann model for polycrystalline solidification with motion of numerous dendrites}},
volume = {245},
year = {2024}
}

@article{Isensee2022,
author = {Isensee, T. and Tourret, D.},
doi = {10.1016/j.actamat.2022.118035},
issn = {13596454},
journal = {Acta Materialia},
pages = {118035},
publisher = {Elsevier Ltd},
title = {{Convective effects on columnar dendritic solidification – A multiscale dendritic needle network study}},
url = {https://doi.org/10.1016/j.actamat.2022.118035},
volume = {234},
year = {2022}
}

@article{Karma1996_EUTECTIC,
author = {Karma, Alain and Sarkissian, Armand},
doi = {10.1007/BF02648952},
issn = {10735623},
journal = {Metallurgical and Materials Transactions A: Physical Metallurgy and Materials Science},
number = {3},
pages = {635--656},
title = {{Morphological instabilities of lamellar eutectics}},
volume = {27},
year = {1996}
}

@article{Choudhury2011_EUTECTIC,
archivePrefix = {arXiv},
arxivId = {1103.4806},
author = {Choudhury, Abhik and Plapp, Mathis and Nestler, Britta},
doi = {10.1103/PhysRevE.83.051608},
eprint = {1103.4806},
issn = {15393755},
journal = {Physical Review E - Statistical, Nonlinear, and Soft Matter Physics},
number = {5},
pages = {1--20},
title = {{Theoretical and numerical study of lamellar eutectic three-phase growth in ternary alloys}},
volume = {83},
year = {2011}
}

@article{Zhang2019_eu,
author = {Zhang, Ang and Du, Jinglian and Guo, Zhipeng and Wang, Qigui and Xiong, Shoumei},
doi = {10.1007/s11663-018-1479-1},
issn = {10735615},
journal = {Metallurgical and Materials Transactions B: Process Metallurgy and Materials Processing Science},
number = {1},
pages = {517--530},
publisher = {Springer US},
title = {{Dependence of Lamellar Eutectic Growth with Convection on Boundary Conditions and Geometric Confinement: A Phase-Field Lattice-Boltzmann Study}},
url = {https://doi.org/10.1007/s11663-018-1479-1},
volume = {50},
year = {2019}
}

@article{Zhang2018_eu,
author = {Zhang, A. and Guo, Z. and Xiong, S. M.},
doi = {10.1103/PhysRevE.97.053302},
issn = {24700053},
journal = {Physical Review E},
number = {5},
pages = {1--13},
pmid = {29906975},
publisher = {American Physical Society},
title = {{Quantitative phase-field lattice-Boltzmann study of lamellar eutectic growth under natural convection}},
volume = {97},
year = {2018}
}
\end{document}